\documentclass[preprint,12pt,authoryear]{elsarticle} 

\usepackage{amssymb}
\usepackage{amsmath}
\usepackage{xcolor}
\usepackage[version=4]{mhchem}
\journal{Micron}

\begin{document}

\begin{frontmatter}



\title{Back-Focal-Plane Imaging and Linear Density Measurements Of Sub-Diffraction Sized Biological Filaments and Particles} %% Article title


\author{Ilya M. Beskin, Jordan Zesch, Emma Hunt, Alexander Weinstein, Ernst-Ludwig Florin} %% Author name

\affiliation{organization={Center for Nonlinear Dynamics, Dept. of Physics, University of Texas at Austin},%Department and Organization 
            city={Austin},
            state={TX},
            country={USA}}

%% Abstract
\begin{abstract}
Biological filaments and their networks are studied to gain a deeper understanding of cell and tissue properties. Imaging of filaments and networks frequently relies on fluorescence microscopy to achieve high-contrast, high-specificity images. However, fluorescence microscopy studies of filament mechanical properties are hindered by phototoxicity, fluorophore induced changes in mechanical properties, and the difficulty of precise local filament thickness measurements. High-contrast label-free methods are needed to visualize filaments under physiological conditions without fluorescence. Higher contrast can be achieved by measuring the transmitted light intensity distribution. This method, known as differential phase contrast, has been implemented in electron and optical microscopy. Similarly, optical tweezers frequently use back-focal-plane detection to track the position of single, trapped, sub-diffraction sized particles with sub-nanometer precision and MHz bandwidth. Here, this method of single particle tracking is extended to visualizing more complex objects such as filaments. We demonstrate that back-focal-plane-detection-equipped optical tweezers can be used for high contrast microscopy by imaging single collagen fibrils and microtubules. By modeling filaments as a line of individual scattering particles, local linear density and thickness is quantified. The sample-orientation-dependent detector response for filaments can be used for a unique background subtraction method. This is demonstrated by removing the protein aggregate background from surface-bound microtubule images. We show that measurements can be made far from the coverslip, making this an excellent tool for studying the link between the structure and mechanics of filaments and filament networks. Optical tweezer setups with back-focal-plane detection can implement this method without significant optical modifications.
\end{abstract}

\begin{graphicalabstract}
\includegraphics[width=13cm]{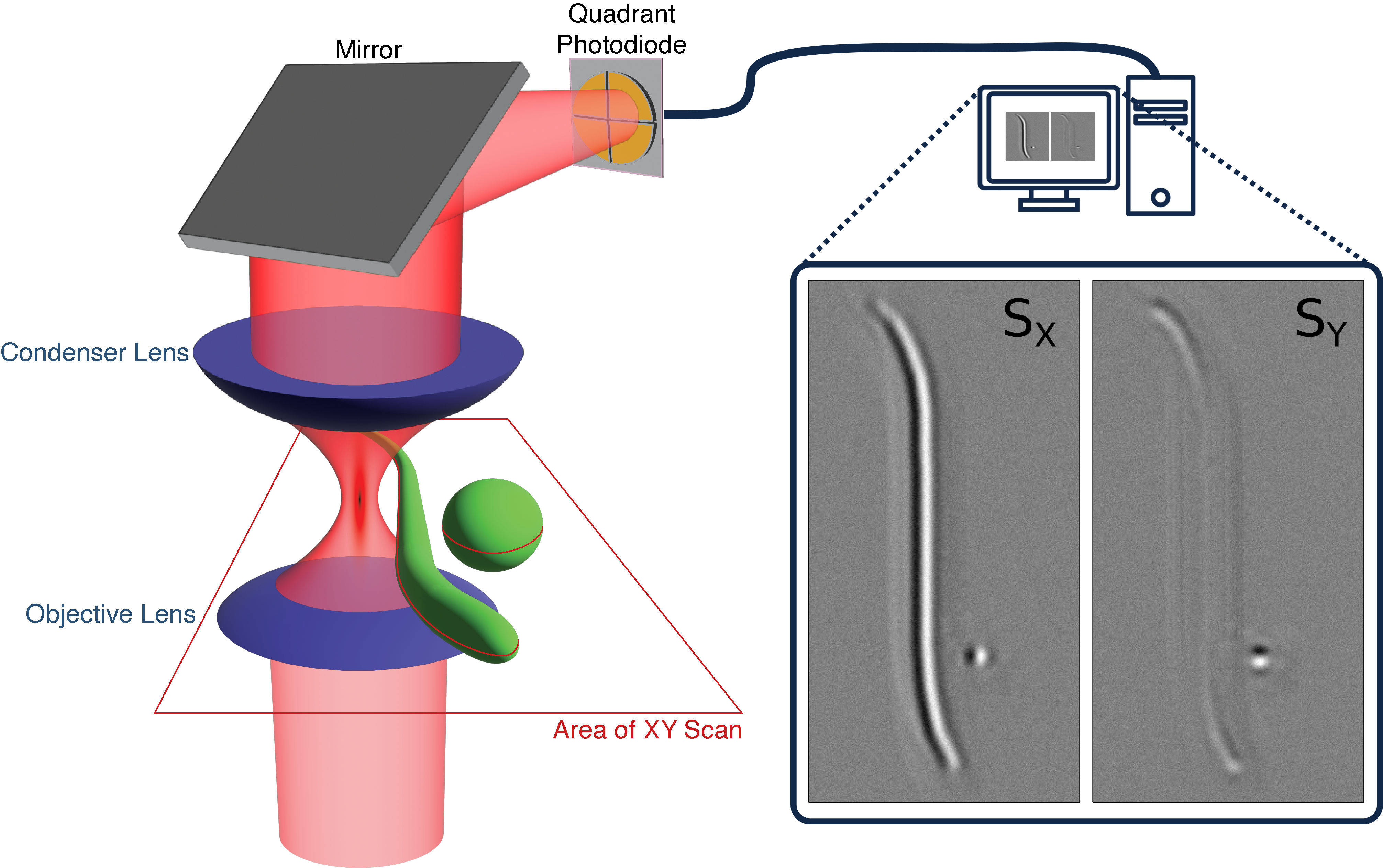}
\end{graphicalabstract}

\begin{highlights}
\item Optical tweezer and differential phase contrast based quantitative microscopy method
\item Measurement mass distributions of filaments and filament diameters
\item Quantitative microscopy method tested on collagen $\alpha$-ends and microtubules
\item Background subtraction algorithm for uniform filaments
\end{highlights}

%% Keywords

\end{frontmatter}




\section{Introduction}

Biological filaments such as collagen and microtubules are responsible for cellular and tissue structure, mechanical properties, and are vital for many cellular processes \citep{CollagenIRev_Biomedicines2022, MTTransAndDistOrganelles_Gelfand_CSHPBiol2017}. For instance, microtubules have been imaged to study their polymerization and de-polymerization to further our understanding of cell division \citep{MTCatAndRescRev_MZanic_JHow_CellArch2013,MTPolym_LukeRice_NatComm2025}. Collagen scaffold deformations caused by a cell have been imaged to explain the dynamics of cell motility \citep{Doyle_DeformECMbyCell_DevCell2021}. Therefore, significant interest exists in imaging these filaments under physiological conditions. Traditional bright-field and dark-field microscopy rely on scattering intensity to visualize objects and often do not achieve the contrast needed for imaging individual filaments and resolving their local thickness. Commonly, fluorescence microscopy is used to visualize filaments with high contrast and specificity \citep{MTFluor_Book_Alieva2020}. However, due to the stochastic nature of labeling and photo-bleaching, measuring accurate local filament thickness with fluorescence microscopy is difficult. Furthermore, the presence of the fluorescence labels may alter the mechanical properties of filaments \citep{MTFluorBreakup_McIntosh_JCellBiol1988, FluorRuinsMT_Traffic2015}. These issues can be avoided by using higher contrast label-free microscopy techniques.

Higher label-free contrast can be achieved by probing the local variation in the index of refraction in the sample. Schlieren imaging uses the deflection of light caused by a gradient in the refractive index to measure minute effects like density variations in air currents \citep{SchlierenRev_Settles_MeasSciTech2017}. A similar technique initially invented for electron microscopy, called differential phase contrast (DPC), was soon extended to optical microscopy \citep{Stewart_DPC_JOptSocAm76, DPC_HamiltonSheppard_JMicr1984} by measuring the asymmetry of the transmitted light pattern generated by refraction through the sample. As a result, this method measures variations in thickness of the imaged object and heterogeneity in its refractive index in the area illuminated by the scanning focused laser beam \citep{MehtaSheppard_DPC_OptLett2009, DPCandDIC_AmosLaufer_JMicr2003, DPC_LubkZweck_PhysRevA2015}. However, to our knowledge, this method has not been extended to studying sub-diffraction sized filaments.

Also inspired by electron microscopy, a similar optical setup is commonly used for particle tracking in optical tweezers \citep{Gittes98}. With growing use of optical tweezers, this detection method was perfected to allow particle tracking in three dimensions with sub-nanometer precision and MHz temporal resolution \citep{PralleEL99, Rohrbach02}. Further modifications of this technique pushed the sampling rate to 100MHz \citep{ELRaizen2011}. This method is known as back focal plane (BFP) detection \citep{OptTweezVolpe}. Both DPC and optical-tweezer BFP detection work through measuring the light intensity distribution in the BFP of the condenser lens. However, while DPC contrast is interpreted as refraction of the beam generated by inhomogeneous phase delays in the sample, optical-tweezer BFP contrast is modeled through computing the interference pattern between an unscattered beam and the light scattered by the small trapped particle \citep{Gittes98, OptTweezVolpe}.

\begin{figure}[htbp]
\centering\includegraphics[width=13cm]{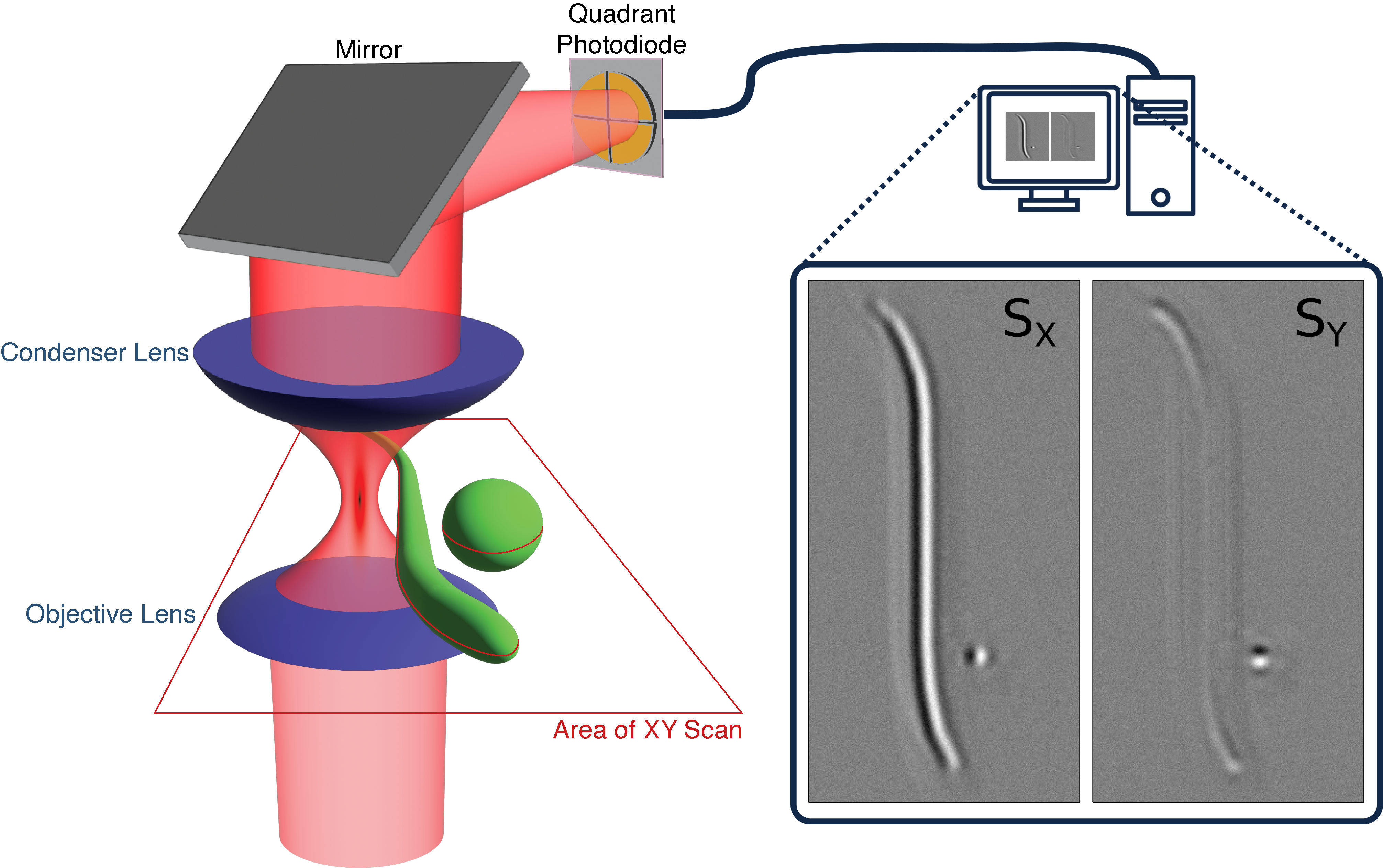}
\caption{Simplified illustration of BFP microscopy method used in this study. An example of imaging a filament and a neighboring particle is shown along with a simulated readout of the $S_x$ and $S_y$ signals that would be observed with this sample geometry. During imaging, the sample is translated through the laser beam's focus. At each sample position, the quadrant photodiode (QPD) signals are recorded as the corresponding pixel value in the QPD response images. The orientation-dependence of the $S_x$ and $S_y$ responses is visible on the computer readout.}
\label{MicrCart}
\end{figure}

Here, we introduce a laser scanning microscopy technique that uses BFP detection for label-free imaging of biological filaments (Fig. \ref{MicrCart}). Analogously to optical tweezer BFP detection, we interpret the resulting contrast as being generated by the interference of the unscattered beam and the light scattered by these filaments. This allows us to accurately measure local linear density and radii of microtubules and collagen fibrils.

A similar imaging technique was used for locating filaments in three dimensions within a filament network \citep{LissekBiorX2018}, but structural information such as filament thickness was not extracted from the detector signal. The ability to extract local diameter variations using the approach presented here is essential for a deeper studies of individual filament and filament network mechanical properties.

In this work, we extract quantitative structural information from imaged filaments by modeling them as a line or Rayleigh scatterers. Each Rayleigh scatterer additively contributes to the signal in each pixel of the final image as long as the intensity of scattered light is significantly smaller than the intensity of the incoming laser beam (discussed further in suppl. sec. S2). We demonstrate the validity of this model by imaging groups of close-by particles in various configurations, by measuring the changing radius at tapered $\alpha-$tips of individual collagen fibrils, and the linear density of microtubules.

The BFP microscopy method presented here is capable of measuring anywhere inside the sample volume, making it optimal for studying systems like biological filament networks. Finally, we make use of the sample orientation-dependent detector response to drastically reduce the background in images of uniform elongated objects.  We demonstrate the efficiency of this method by removing the background in images of individual microtubules that originates from point-like protein aggregates. We achieve a signal-to-background ratio (SBR) comparable to the best reported for label-free microscopy techniques \citep{IntReflMicrSBR_ESchaf_JHow_2018,iScat_SBR_Kukura2016}.

\section{Back Focal Plane Detection in Optical Tweezer Systems}
BFP detection can be used to achieve nanometer-scale particle position detection in optical tweezer systems. This works through measuring the interference between unscattered light and light scattered by the imaged structure. To create this interference pattern, a laser beam is expanded to overfill the back aperture of a high-numerical aperture microscope objective lens. After passing through the lens, the light converges to a diffraction-limited spot in the liquid sample. This focus is the location of the optical trap, assuming that the light intensity is sufficiently high to produce a stable trapping potential.

If a small particle is located near the focal point of the laser, a small portion of the laser light is scattered by the particle. The interference of the scattered and unscattered light allows us to extract the position of the particle relative to the focus\citep{Gittes98,PralleEL99}. Both the scattered light and the unscattered beam are then collected by a microscope condenser lens. After the condenser lens, the laser light passes through a pair of convergent lenses that allows us to insert filters into the detection beam path. Finally, the light is projected onto a quadrant photodiode (QPD) in a plane conjugate to the BFP of the condenser lens. The QPD, frequently used to measure laser beam pointing fluctuations or asymmetries in a laser beam, consists of four monolithic photodiodes, one in each quadrant, separated by small gaps. A differential preamplifier processes the QPD signals providing three readings: the intensity difference between the right and left photodiodes, the difference between the top and bottom photodiodes, and the total intensity on all four photodiodes (see suppl. sec. S1). These three signals are referred to respectively as the $S_x$, $S_y$, and $S_z$ signals. In this paper, we will limit our examination to the $S_x$ and $S_y$ signals. Being differential signals, they are less sensitive to laser power noise.

While imaging loosely bound objects, the trapping forces of optical tweezers could cause shifts in the object's position. To avoid this, we attenuate our laser with a neutral density filter by 95\% relative to the power typically used for trapping particles. This low intensity ($\approx 0.13\ mW$ at the beam waist) the  applied trapping forces are negligible. More specifically, we apply a trapping potential small enough to not affect the magnitude of thermal fluctuations experienced by the imaged structures (see supl. sec. S4). As our sample preparation protocols lead to sufficiently strong attachment to avoid significant motion blur, the weak trapping potential similarly does not alter the images. Furthermore, thermal heating from optical trapping with a $1064\ nm$ laser in water has been measured by Erwin J.G. Peterman et al. \citep{PetermanGittesSchmidt_Heating_BiophysJ2003}. Given our laser power and their measurements, we estimate the effect of thermal heating to be less than $0.01\ K$ at the laser focus. Despite the low laser power, this intensity is sufficient to create high signal-to-noise images of nanometer scale objects.

\subsection{Imaging Single Particles}
To understand the QPD response to a single point-like particle, we first have to calculate the interference pattern at the QPD (see suppl. sec. S1), i.e. find the sum of the electric fields of the scattered and unscattered light at the QPD:
\begin{align}
    I = \dfrac{\epsilon_s c_s}{2}\left[(E_{beam}+E_{scat})(E_{beam}^*+E_{scat}^*)\right]
\end{align}
where $E_{beam}$ is the electric field of the unscattered beam, $E_{scat}$ is the electric field of the light scattered by the particle, $*$ represents the complex conjugate, and $\epsilon_s$ and $c_s$ represent the permittivity and speed of light in the solution filling the sample. For small particles, we can make the approximation that the field of the unscattered beam, $E_{beam}$, is equal to the field of the beam that doesn't encounter any particles in its path (see suppl. sec. S2). Since we are interested in differential QPD signals generated by subtracting one side of the QPD from the other side of the QPD, any part of our beam that is axially symmetric around the optical axis will not contribute to the QPD signal. As we center our QPD on the laser beam at the start of each experiment, a beam that doesn't encounter any particles will not contribute to the $S_x$ or $S_y$ signals. Thus, it is sufficient to calculate the perturbation to the intensity:
\begin{align}
    \delta I &= \dfrac{\epsilon_s c_s}{2}\left[(E_{beam}+E_{scat})(E_{beam}^*+E_{scat}^*) - E_{beam}E_{beam}^*\right]    \label{eqn:ReEE}\\
    &= \dfrac{\epsilon_s c_s}{2}\left[E_{scat}E_{beam}^* + E_{scat}^*E_{beam} +E_{scat}E_{scat}^* \right] \\
    &\approx \epsilon_s c_s \mathbb{R}e\Big\{E_{scat}E_{beam}^*\Big\}
\end{align}

Note that we consider the $E_{scat}E_{scat}^*$ term negligible in the small particle limit, where the scattered electric field is much smaller than the electric field of the incoming beam. In order to compute the signals $S_x$ and $S_y$, we take the Rayleigh approximation and treat the scattering particle as a dipole (see suppl. sec. S2 and S3). A more refined treatment needed for larger particles of varied shapes requires a Mie scattering calculation, which can be performed using the T-Matrix method \citep{TMatrix_Waterman1965}. The application of the T-Matrix method to the optical tweezer detection path is described in \citep{OptTweezVolpe}. We calculate the interference patterns for a variety of particle sizes using code based on the ``OTS-the Optical Tweezers Software'', a part of \citep{OptTweezVolpe}.

In order to compare our microscopy images of single particles with predictions from Rayleigh and Mie scattering computations, we image five $110\ nm$ diameter polystyrene beads attached to a coverslip surface. After finding the correct sample height (see suppl. sec. S5), multiple scans in the $XY$ plane of each particle are taken. The scans are aligned with each other and averaged to increase the signal-to-noise ratio and to average over variations in particle diameter. By choosing beads with a diameter of $110\ nm$, we ensure that the intensity of the scattered light is much smaller than the intensity of the unscattered light (i.e. $E_{scat}<<E_{beam}$). Also, the particle diameter is much smaller than the beam waist ensuring that all points in the particle experience a uniform electric field from the illuminating laser; an assumption made in predicting the detector response with the Rayleigh scattering computation described in suppl. sec. S2. The beam waist is defined as the radius at which the light intensity in the focal plane decreases by $1/e^2$. In our system, the beam waist is measured to be $w_0\approx400\ nm$. Therefore, $110\ nm$ diameter particles are a good compromise between acting as point-like objects and generating a large enough signal-to-background ratio.

\begin{figure}[htbp]
\centering\includegraphics[width=13cm]{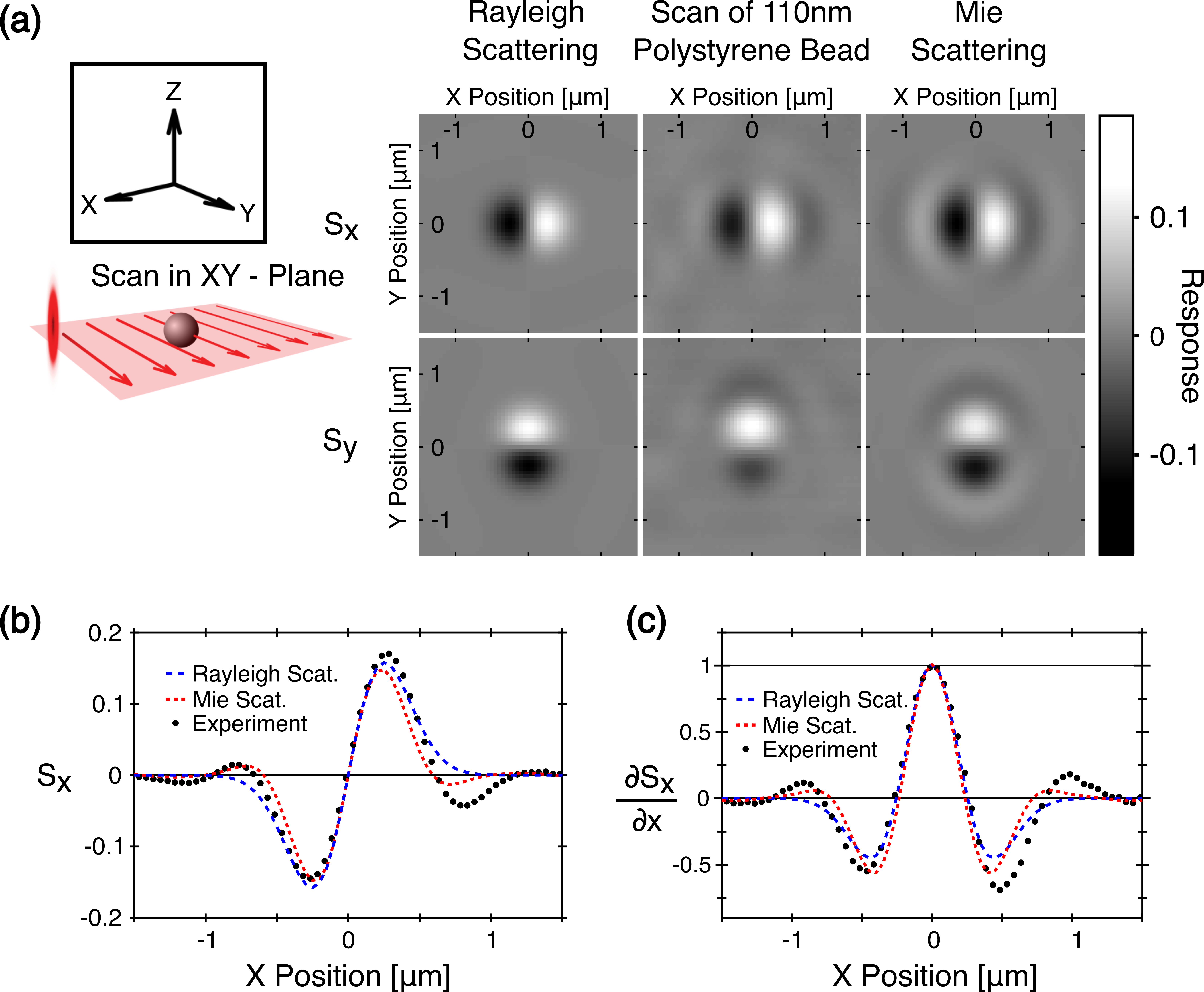}
\caption{Experimental and simulated detector response to a single particle. Average $S_x$ and $S_y$ taken from scans of $110\ nm$ diameter polystyrene beads are compared to a computed Rayleigh scattering response and a computed Mie scattering response. \textbf{(a)} Computed responses in the $XY$ plane compared to experimental scan of a single bead. Mie scattering computed response uses a $110\ nm$ bead. Rayleigh scattering response was computed for a single dipole rescaled to represent a $110\ nm$ bead (see suppl. sec. S1). \textbf{(b)} Line profiles of the $S_x$ computed and experimental responses taken along $y=0$. \textbf{(c)} Sensitivity of $S_x$ for computed and experimental responses from part (b). Note, since the experimental response amplitude depends on specific instrument settings, all responses in (c) have been rescaled such that the maximum $S_x$ sensitivity, $(\partial S_x / \partial x )|_{x=0}=1$.}
\label{SimVsRealResp}
\end{figure}

Fig. \ref{SimVsRealResp}a shows the average $S_x$ and $S_y$ response for a $110\ nm$ polystyrene bead next to the Rayleigh and Mie scattering computational results. When comparing the area of the non-zero part of the experimentally measured and simulated response curves (Fig. \ref{SimVsRealResp}a : middle), we find that it is modeled well by both the Rayleigh and Mie scattering computations (Fig. \ref{SimVsRealResp}a : left and right). This indicates that the models correctly approximate the QPD $S_x$ and $S_y$ responses. Both the simulated $S_x$ responses and the experimental $S_x$ response display symmetry with respect to reflection across the $x-$axis and antisymmetry with respect to reflection across the $y-$axis leading to the characteristic vertical ``black-white transition''.  All three $S_y$ responses display symmetry with respect to reflection across the $y-$axis and antisymmetry with respect to reflection across the $x-$axis creating a horizontal ``black-white transition''. A slight elongation of the $S_y$ response relative to the $S_x$ response is caused by the laser polarization. Line profiles of the $S_x$ signal along the $y=0$ line are shown in Fig. \ref{SimVsRealResp}b. Here, we find good agreement in the width and general shape of the $S_x$ signals. The more prominent ``overshoot'' seen in the experimental data and Mie scattering computation around $x-$position $=\pm0.7\ \mu m$ is not shown correctly in the Rayleigh computation, as we use the Gaussian beam approximation for the unscattered beam. From the $S_y$ responses in Fig. \ref{SimVsRealResp}a, one can see that the intensity of the experimental signal is lower than the intensity from both computations, and the experimental response is not fully antisymmetric. This is due to a small optical misalignment along the $y-$direction. Despite this difference, we see good agreement between the experimental and computational responses. Also, the imaging method presented in following sections is robust against such instrument-specific variations.

\subsection{Signal Scaling with Particle Size} \label{sec:beadSizeResp}
\begin{figure}[htbp]
\centering\includegraphics[width=12cm]{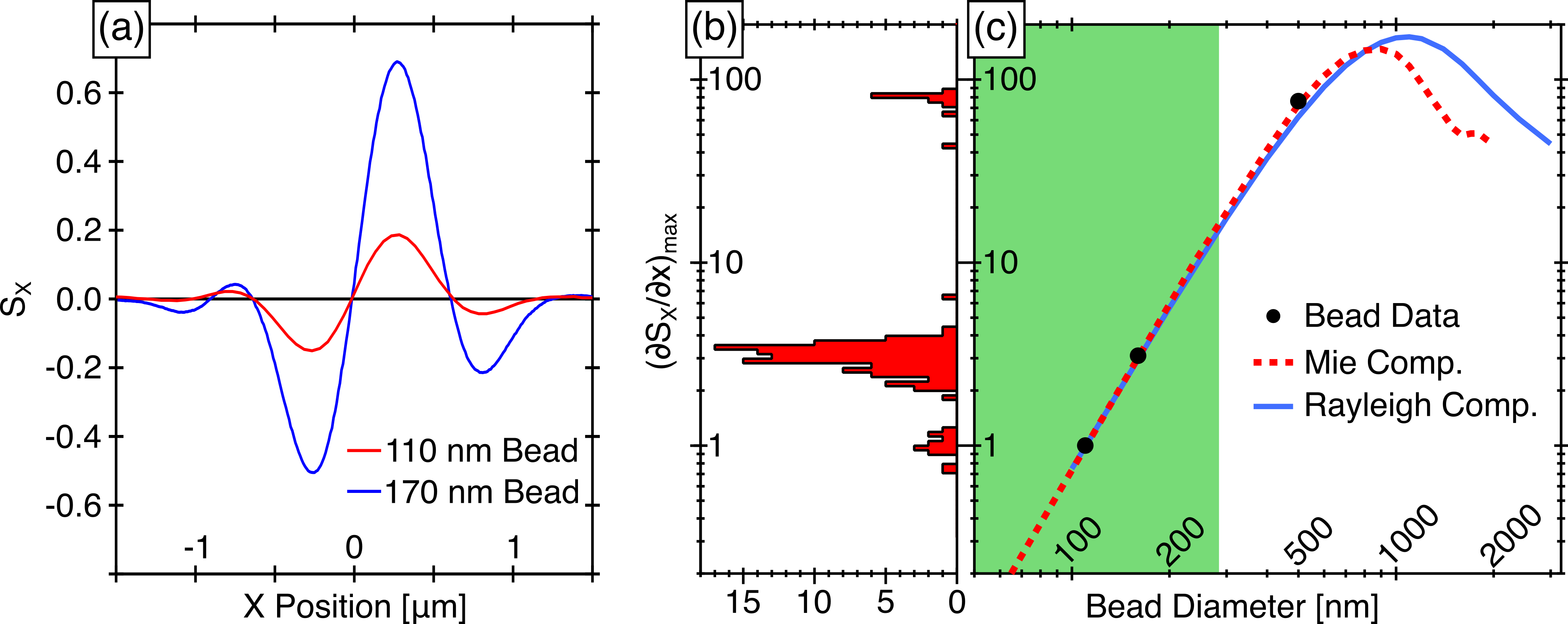}
\caption{Bead response scaling at varying diameters. \textbf{(a)} Line profile of experimental $S_x$ response curve for a $110\ nm$ and $170\ nm$ diameter polystyrene bead. \textbf{(b)} Histogram of bead $\left(\partial S_x/\partial x\right)_{max}$ taken from a single sample of $110\ nm$, $160\ nm$, and $500\ nm$ diameter beads attached to a coverslip surface. The histogram shows each group of beads. \textbf{(c)} Average sensitivity for each bead size plotted along with sensitivities found via Rayleigh and Mie scattering computations. Both computations predict that the maximum sensitivity of beads smaller than $240\ nm$ in diameter (area shown in green) is within $1\%$ of the maximum sensitivity predicted by scaling the signal with bead volume. Error bars representing the error of the mean in $\left(\partial S_x/\partial x\right)_{max}$ values. The error of the mean in bead diameter is smaller than the marker size. Values have been rescaled such that $\left(\partial S_x/\partial x\right)_{max}=1$ for a $110\ nm$ polystyrene bead.}
\label{fiScat}
\end{figure}

Based on our computed Rayleigh scattering response (see suppl. sec. S2), we expect the magnitude of the interference pattern to change proportionally to the volume of the scanned bead while its shape remains constant. Therefore, the $S_x$ and $S_y$ responses should grow proportionally to the radius of the bead cubed for small beads. The experimental responses of $110\ nm$ and $170\ nm$ diameter beads plotted together ( Fig. \ref{fiScat}a ) confirm that the shape of the response doesn't change.

In order to compare the magnitudes of the $S_x$ and $S_y$ responses for different sized beads, we use a measurement referred to as ``maximum sensitivity''. The sensitivity is a measurement of the change in the $S_x$ or $S_y$ response produced by shifting the measured object by a small distance in the $x$ or $y-$direction respectively, i.e. the partial derivative of the response with respect to the associated direction:
\begin{align}
    S_x\textrm{ Sensitivity } \overset{\textrm{defn.}}{=} \  \dfrac{\partial S_x}{\partial x} \qquad \textrm{and} \qquad  S_y\textrm{ Sensitivity } \overset{\textrm{defn.}}{=} \  \dfrac{\partial S_y}{\partial y}
\end{align}

The $S_x$ sensitivity of the $XY$ scans plotted along the $y=0$ line profile is shown in Fig. \ref{SimVsRealResp}c. The maximum value of $\partial S_x/\partial x$ is located at the bead-laser displacement of $x=0$. We denote the maximum $S_x$ or $S_y$ sensitivity value by $\left(\partial S_x/\partial x\right)_{max}$ or $\left(\partial S_y/\partial y\right)_{max}$, respectively. For larger particles, the assumption that all parts of the particle experience the same electric field from the illuminating laser beam is not fulfilled. To test whether the maximum sensitivity scales with volume and to find the particle diameter at which this scaling law breaks, we scan $110nm$, $160nm$, and $500nm$ diameter carboxylated beads attached to a coverslip via poly-l-lysine (PLL) in the same sample. These three bead sizes were chosen as they are made from an identical material with identical fluorescent labeling by the same manufacturer. As the $S_x$ and $S_y$ sensitivities are maximized when the laser focal plane matches the height of the imaged object, we locate the optimal surface height (see suppl. sec. S5) before each scan.

$XY$ scans of over 100 particles are performed and $\left(\partial S_x/\partial x\right)_{max}$ of each bead computed. As the $S_x$ response values depend on instrument parameters such as amplifier gains, we chose to rescale all values by the maximum $S_x$ sensitivity of a $110\ nm$ polystyrene bead. The histogram of sensitivities shows three distinct groupings corresponding to each of the three particle sizes ( Fig. \ref{fiScat}b ). The mean and standard deviation of the maximum sensitivity values for each grouping was computed and plotted versus their average size ( Fig. \ref{fiScat}c ). Mie scattering and Rayleigh scattering computations are performed for a variety of bead sizes. Unlike the computation presented in suppl. sec. S2, the Rayleigh scattering computation performed here takes into account the non-uniform field experienced by larger particles (see suppl. sec. S6).

The results of the Mie scattering computation and the refined Rayleigh scattering computation can be seen in Fig. \ref{fiScat}c. For small bead diameters, the experimental data, as well as both computations, show the expected scaling with particle volume. Specifically, the trend line between the $110\ nm$ and $160\ nm$ diameter particles returns a fit of $(\partial S_x/\partial x)_{max} \propto (diameter)^{3.02\pm0.04}$. Both the Mie and the Rayleigh scattering simulations similarly return a slope of $3$ on the log-log plot for small bead diameters. More specifically, below a bead diameter of $240\ nm$, the maximum sensitivity of the bead, predicted by both models, deviates by less than $1\%$ from the sensitivity predicted by simply scaling the bead maximum sensitivity with volume. Slight deviations from the expected cubic growth become visible at bead diameters above $240\ nm$. This is expected as our beam waist is $\approx 400\ nm$ and larger particles start to experience a less uniform electric field. As one approaches $1\mu m$ sized particles, the Mie scattering simulation shows a deviation from the Rayleigh scattering simulation. In this regime, the bead starts scattering a larger quantity of light. This invalidates the assumption made in the Rayleigh scattering computation that scattered light does not have multiple interactions with the particle. However, for objects with diameters below $240\ nm$, the maximum sensitivity of a signal is linearly proportional to the object's mass.

Assuming that the scanned objects have a known index of refraction and density and the optical tweezer microscope has been calibrated by scanning a well-characterized reference bead, the scaling law can be used to determine the mass of a variety of other objects. Mass distributions of extended objects are discussed in following sections. In order to compute the volume and mass of a point-like object, the following steps are performed. First, using the relationship of refractive index to scattering intensity \citep{Gittes98}, we are able to compute the maximum sensitivity that would be generated by a reference bead of the same volume, $V_{rb}$, but with the refractive index of the measured object. To do this, the maximum sensitivity from the reference bead (in our case, a $110\ nm$ diameter polystyrene bead) is multiplied by the correction factor $n_{corr}$:
\begin{align}
    n_{corr} = \left(\dfrac{(n_{obj}/n_s)^2-1}{(n_{obj}/n_s)^2+2}\right)\Big/\left(\dfrac{(n_{rb}/n_s)^2-1}{(n_{rb}/n_s)^2+2}\right) \label{eqn:nCor}
\end{align}
where $n_{obj}$, $n_s$, and $n_{rb}$ are the refractive indices of the scanned object, surrounding sample fluid, and the reference bead, respectively. As volume scales with the measured maximum sensitivity, we are now able to compute the volume of the measured object $V_{obj}$:
\begin{align}
    \dfrac{V_{obj}}{V_{rb}} = \dfrac{(\partial S_x/\partial x)_{max}^{obj}}{n_{corr} \times(\partial S_x/\partial x)_{max}^{rb}}
\end{align}
where $(\partial S_x/\partial x)_{max}^{obj}$ and $(\partial S_x/\partial x)_{max}^{rb}$ are the maximum sensitivities of the scanned object and the reference bead, respectively. In order to find the mass of the imaged object, $m_{obj}$, the volumes are converted to masses using the known densities:
\begin{align}
    \dfrac{m_{obj}}{m_{rb}} = \dfrac{\rho_{rb}}{\rho_{obj}}\times\dfrac{(\partial S_x/\partial x)_{max}^{obj}}{n_{corr} \times(\partial S_x/\partial x)_{max}^{rb}}
\end{align}

As the mass of the scanned object is determined by the sensitivity, we use the uncertainty in the fit to the $S_x$ response to find the mass resolution limit. We can resolve the mass of polystyrene particles up to $950\ kDa$. For a more biologically relevant example, we can also estimate the mass resolution for globular proteins to be $1.1\ MDa$. We used Eqn. \ref{eqn:nCor} to take into account the index of refraction and specific volume of globular proteins ($n=1.587\pm0.005$ and $0.745\pm0.009\ cm^3/g$  with little variation \citep{iScatProtMass_Science2018}). The index of refraction and density of polystyrene are known to be $n=1.572$ at $1064\ nm$ \citep{nPolysty_ActaPhysPolonicaA2009} and $1.05 g/cm^3$ (from manufacturer specifications). Currently, electronic noise, laser noise, vibrations, and inhomogeneity in the sample limit the mass resolution of our instrument. (see a discussion of laser power noise, pointing noise and vibrations in suppl. sec. S8). Except for sample inhomogeneity, the sources of noise can be mitigated with sufficient averaging.

\section{Imaging Multiple Particles}

\subsection{Imaging Multiple Beads}

\noindent In order to use optical tweezers for microscopy, we must understand the images produced by extended objects. A simple extended object consists of multiple nearby particles. If imaging two particles whose scattered electric fields are represented by $E_{p1}$ and $E_{p2}$, we can calculate the detector signal (analogously to the computation in Eqn. \ref{eqn:ReEE}):
\begin{align}
    \delta I &= \dfrac{\epsilon_s c_s}{2}\left[ \left|E_{beam} + E_{p1} + E_{p2}\right|^2-|E_{beam}|^2\right]\\
    %&= \dfrac{\epsilon_s c_s}{2} \left[ (E_{beam}+E_{p1}+E_{p2})(E_{beam}^*+E_{p1}^*+E_{p2}^*) - E_{beam}E_{beam}^* \right] \\
    %\begin{split}
    %    &= \dfrac{\epsilon_s c_s}{2} \Big[ E_{beam}E_{p1}^* + E_{beam}E_{p2}^* + E_{beam}^* E_{p1} + E_{beam}^* E_{p2} \\
    %    &\hspace{2cm} + E_{p1}E_{p2}^* + E_{p1}^*E_{p2} + E_{p1}E_{p1}^* + E_{p2}E_{p2}^* \Big]
    %\end{split}\\
    &\approx \epsilon_s c_s \left[ \mathbb{R}e\Big\{E_{p1}E_{beam}^*\Big\} + \mathbb{R}e\Big\{E_{p2}E_{beam}^*\Big\} \right]
    \label{eqn:ReEEpReEE}
\end{align}
Here we again neglected the products of two scattering terms, $E_{p1}$ and $E_{p2}$, as the scattering intensity is tiny compared to the intensity of the unscattered beam. Thus, assuming that the scattering intensities are small compared to the unscattered beam, the interference pattern on the principal surface of the condenser for two particles can be represented as the sum of the interference patterns of the respective particles. As a result, the $S_x$, $S_y$, and $S_z$ signals generated by two particles are also well approximated by the sum of the respective signals that would have been generated by each of these particles individually. This result holds for particles of different sizes and orientations relative to the axes of the QPD. The calculation can be repeated for more than two particles. It shows that our approximation holds largely true for and configuration of multiple small particles, which allows a simple quantitative analysis of images collected through BFP differential microscopy.

\begin{figure}[htbp]
\centering\includegraphics[width=10cm]{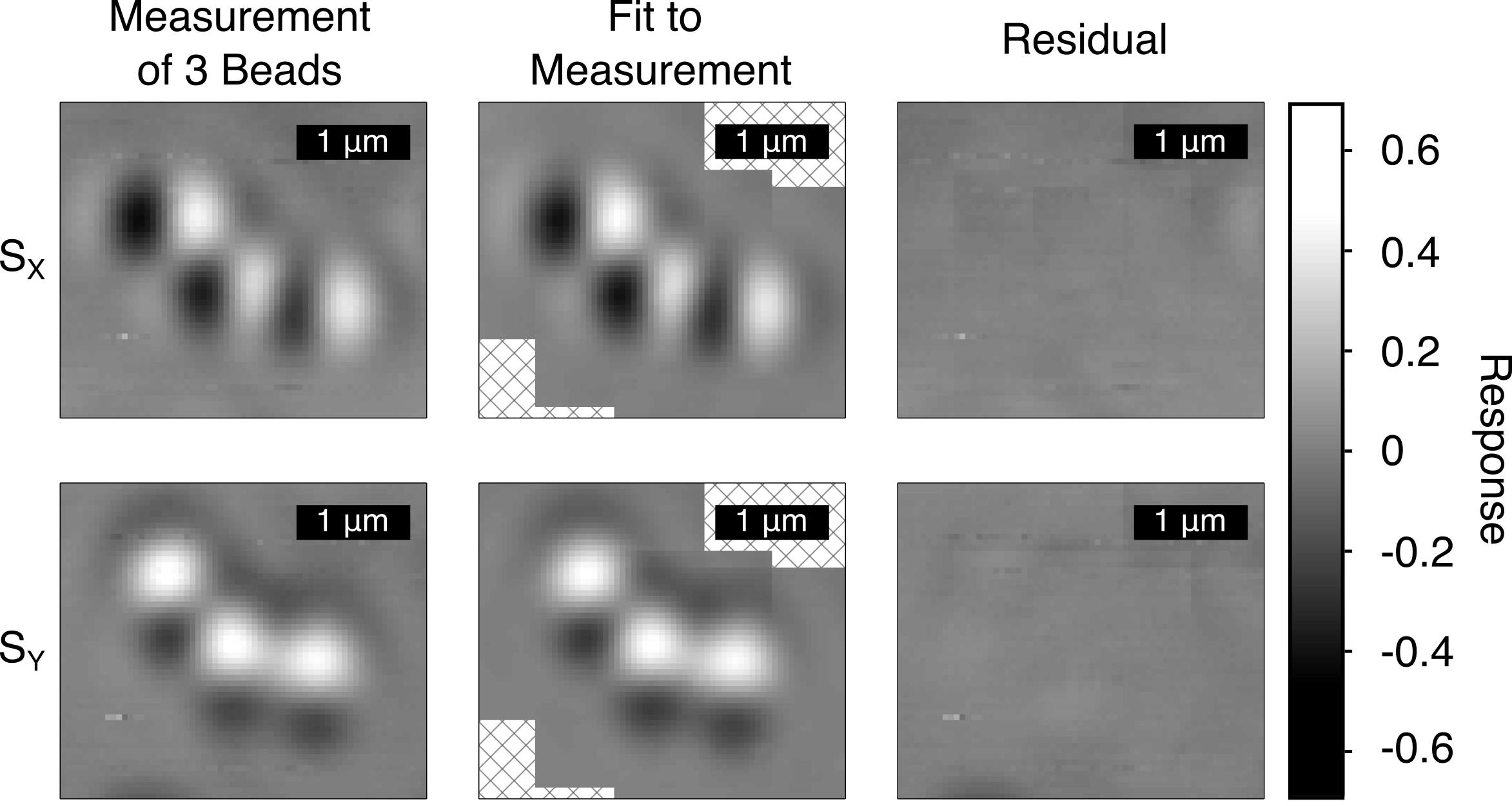}
\caption{Experimental response for multiple nearby beads compared to the sum of response from single beads. In the left column, experimental $S_x$ and $S_y$ responses of three close $170nm$ beads attached to a coverslip are shown. The middle column shows the fit function consisting of the weighted sum of three shifted single bead response curves. The right column shows the residuals of the fits. Responses in this figure are rescaled such that $\left(\partial S_x/\partial x\right)_{max}=1$ for a $110\ nm$ polystyrene bead.}
\label{3Bead}
\end{figure}

To experimentally verify that images of extended objects can be understood as the sum of images of individual particles, $170\ nm$ carboxylated polystyrene beads are placed in a sample chamber filled with water and allowed to settle at random on a PLL coated coverslip. A imaged region containing three particles with overlapping response curves is shown in Fig. \ref{3Bead} : left. In order to show that the resulting three-particle response is approximated by the sum of three single-particle responses, we write a custom fit function which uses the experimentally determined $S_x$ and $S_y$ responses (Fig. \ref{SimVsRealResp}a : middle ) of a $110\ nm$ diameter particle as a template. We perform a least-squares fit with a fit function $f(C_1,\dots,y_n)$ consisting of a weighted sum of three copies of the template. The weight and $x$ and $y$ shift of each template in the sum are free fitting parameters. This fit function, generalized for $n$ beads of potentially different sizes, is presented in Eqn. \ref{eqn:multiBeadFit}.
\begin{align}
    \begin{split}
        f(C_1,x_1,y_1&, \dots ,C_n,x_n,y_n) = \\ &=\begin{pmatrix}C_1S_x(x-x_1,y-y_1) + \dots + C_nS_x(x-x_n,y-y_n) \\
        C_1S_y(x-x_1,y-y_1) + \dots + C_nS_y(x-x_n,y-y_n)
        \end{pmatrix}
    \end{split}
    \label{eqn:multiBeadFit}
\end{align}
The $S_x$ and $S_y$ responses for the three beads (Fig. \ref{3Bead} : left) are visually similar to the fit constructed using the sum of three single bead responses (Fig. \ref{3Bead} : middle). We can confirm that the responses indeed match by looking at the residual of the fit (Fig. \ref{3Bead} : right). The residual is dominated by pixel-to-pixel noise indicating that the $S_x$ and $S_y$ responses from multiple beads can be represented by the sum of individual responses. By using an experimental $XY$ scan of a small bead as a reference, instrument specific parameters such as an objective lens specific point spread function, are accounted for in the fit. As a result, the residuals of the $S_y$ fit are as small as the residuals of the $S_x$ fit despite the small optical misalignment seen in the $S_y$ response in Fig. \ref{SimVsRealResp}a. This underlines that our method is robust against instrument-to-instrument variation as long as a calibration scan of a single point-like object is used.

In order to test that this remains true for beads at different positions and orientations relative to one another, we placed $170\ nm$ carboxylated polystyrene beads into a sample filled with phosphate-buffer saline solution (PBS). This buffer, frequently used to mimic the ion concentration and pH of the extracellular environment, reduces the electro-static repulsion between particles allowing them to be located closer to each other or even coalesce as they settle on the coverslip. An area with a high density of particles was scanned (see suppl. sec. S7) to confirm the results seen in Fig. \ref{3Bead}. The fit captures the majority of the signal (Fig. S6a). The residuals show a structure-correlated pattern remains that is an order of magnitude smaller than the original image. Due to the large concentration of particles located in close proximity and potentially stacked on top of each other, the magnitude of the scattered light grows, challenging the approximation that $E_{scat} << E_{beam}$ (i.e. the product of the terms $E_{p1}$ and $E_{p2}$ can no longer be fully neglected). However, the prediction of Eqn. \ref{eqn:ReEEpReEE} that the sum of single bead responses is a good approximation for the response to multiple beads, holds regardless of relative bead orientation or spacing between beads, even with such a high density of relatively large particles. In addition to reproducing the image, the weights and locations found using the template fit function (Eqn. \ref{eqn:multiBeadFit}) are expected to give the mass distribution in the imaged sample. To verify that the positions of the particles are correctly identified, we compare them with positions found using fluorescence microscopy (Fig. S6b). On average, the distance between the positions identified using the template fit method and fluorescence is $50\ nm$. Both fluorescence intensity and the weight of the template fit (as shown in section \ref{sec:beadSizeResp}) scale with the volume of the bead. We confirm this by graphing the relationship between fluorescence intensity and template fit weight (Fig. S6c). The result shows the expected linear relationship with the larger particles approximately twice the size of the smaller particles. This is likely the result of particles coalescing in solution before settling on the coverslip surface. A distinct separation between the single and double particles is not seen due to the large variation in single-particle size (individual fluorescent beads have an error in their diameter of $10\%$ from the mean diameter). In conclusion, the locations and masses (i.e. the mass distribution) of extended objects can be accurately identified from our BFP microscopy images as long as the intensity of scattered light is significantly smaller than the beam intensity.

\subsection{Imaging Filaments} \label{sec:ImFil}
Optical tweezers are frequently used to study filaments by either conducting measurements near filaments \citep{SvobodaSchmidtBlock1993, Supertwist_Schaffer2018}, within network structures \citep{ viscoElastMicrorheol_WirtzKuo_PRL1997, TNI_TischerEL_APL2001}, or probing the filaments directly \citep{FelgnerSchliwa_BentingMTwithOT_JCellSci1996,LissekBiorX2018}. In order to model an image of a filament, we can envision the a filament as a row of beads (``bead model''). Thus, imaging a filament is just a specific case of imaging multiple beads. More specifically, one can envision a long cylindrical object of radius $r$ as a being subdivided into cylindrical sections of radius $r$ and height $h$. As the shape of the QPD response of an object smaller than the beam waist does not depend on the object's shape, only on the volume, the response of each cylindrical section will be given by the response of a spherical bead of radius $r$ at that location multiplied by the correction factor:
\begin{align}
    \dfrac{\mathrm{Volume\ of\ Cylinder}}{\mathrm{Volume\ of\ Sphere}} = \dfrac{3h}{4r}
\end{align}

This allows us to place each bead in our ``bead model'' arbitrarily close to each other along the central axis of the filament. To avoid seeing the periodicity of the bead spacing in our simulated ``bead model'' QPD response, we used a spacing significantly smaller than the beam waist. In the data presented here, a center to center bead spacing of $50\ nm$ has been used.

\begin{figure}[htbp]
\centering\includegraphics[width=10cm]{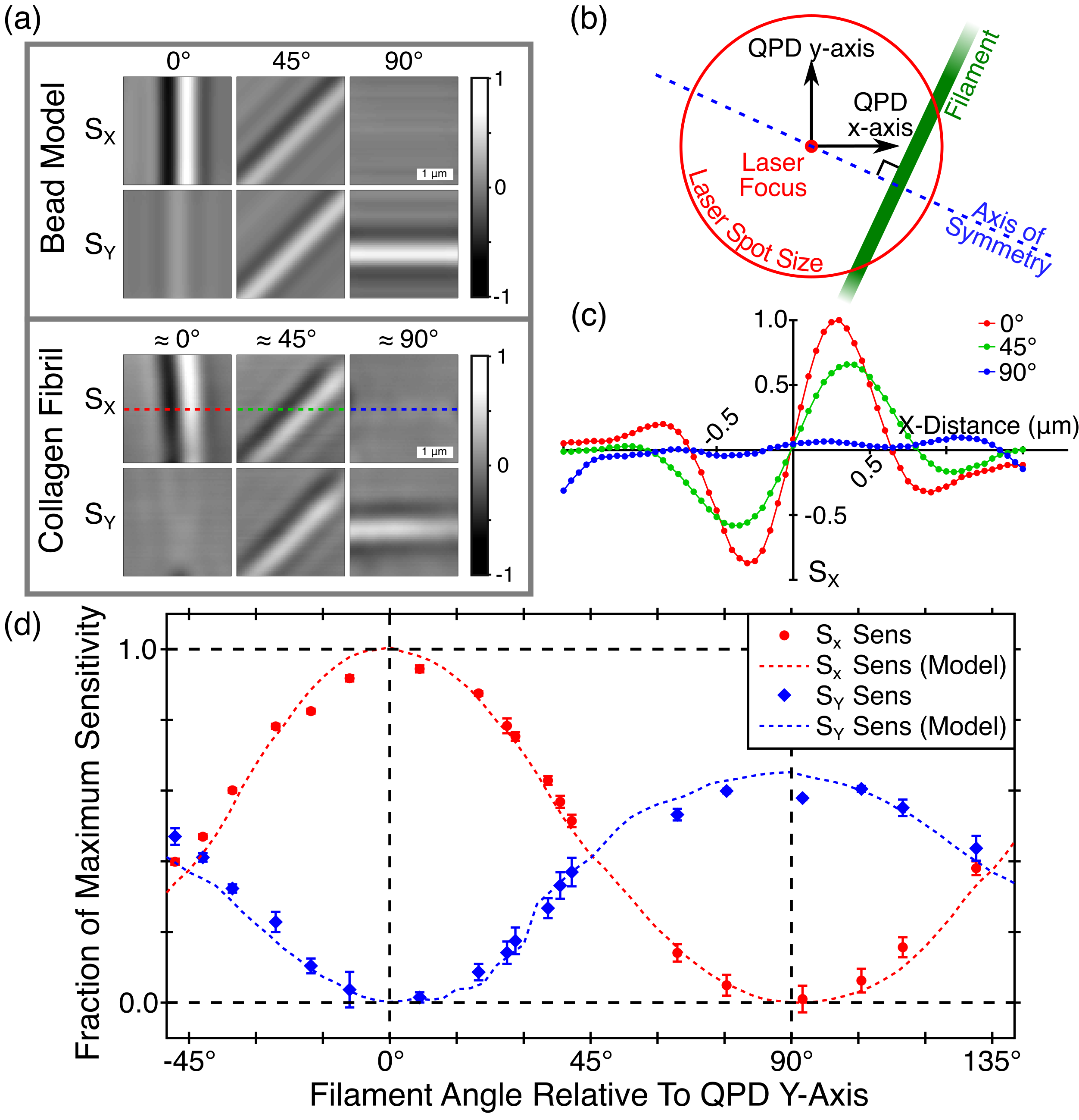}
\caption{The response to a filament compared to the response generated by the filament `bead model'. \textbf{(a)} Scans computed using the bead model of a filament (top) and experimental data  of a collagen fibril (bottom) at different angles relative to the QPD axis. \textbf{(b)} Illustration of the symmetry of the experimental geometry that leads to orientation dependent response for filament-like objects. \textbf{(c)} Line profiles of the $S_x$ responses for the collagen fibril are shown in fig. \ref{Fibrils}b. The location of where the line profile was taken is marked with the associated color. For comparison, plots in parts b and c have been rescaled such that the maximum and minimum $S_x$ response values are $1$ and $-1$ respectively. \textbf{(d)} Measured $S_x$ and $S_y$ max. sensitivity for a region on a collagen fibril plotted along with the maximum sensitivity predicted by the ``bead model''. Plot in part d has been rescaled such that the max. $S_x$ sensitivity is $1$ for a filament aligned with the QPD $y-$axis.}
\label{Fibrils}
\end{figure}

As before, taking the single bead responses (Fig. \ref{SimVsRealResp}a: middle) as a template, we construct the expected $S_x$ and $S_y$ responses of the ``filament bead model'' for filaments at $0^\circ$, $45^\circ$, and $90^\circ$ relative to the QPD $y-$axis (Fig. \ref{Fibrils}a:top ). Note that the response in the $S_x$ and $S_y$ channel depends on the relative orientation of the filament axis to the respective channel. This can be explained by examining the symmetry in the experimental setup and sample geometry (Fig. \ref{Fibrils}b). If a uniform section of a filament is illuminated by the laser, the perpendicular drawn from the laser focus to the filament will serve as an axis of symmetry. The interference pattern generated by this geometry will exhibit the same symmetry. Recall that the QPD  $S_x$ and $S_y$ responses measure the difference in intensity between the right and left, and top and bottom sides, respectively. If this axis of symmetry aligns with the QPD $y-$axis, the interference pattern will be symmetric on the left and right, causing the $S_x$ signal to read zero. Similarly, if the axis of symmetry aligns with the QPD $x-$axis, the $S_y$ signal will read zero.

To compare images of filaments with predictions from the bead model, we image a uniform collagen fibril at different orientation angles relative to the QPD $y-$axis. $500\ nm$ polystyrene carboxylated beads are added to the sample in a low concentration as a position reference. Isolated fibrils are first localized by polarization microscopy. In the following step, the fibril height was optimized using the protocol described in suppl. sec. S5 and a $XY$ scan is taken. After a uniform section of the fibril was identified, we rotate the sample and use the fluorescent beads to find the exact same location on the fibril. The process is repeated for 19 different angles spanning $\approx 180^\circ$. Fig. \ref{Fibrils} (b: bottom) shows three scans at approximately $0^\circ$, $45^\circ$, and $90^\circ$. The $S_x$ and $S_y$ signals show the angle dependence predicted by the filament bead model.

To quantify the angular dependence of the detector signals, multiple line profiles are taken across each scan (Fig. \ref{Fibrils}c). The maximum $S_x$ sensitivity decreases with an increase in the angle of the filament relative to the QPD $y-$axis. Each row of the $S_x$ response and column of the $S_y$ response provide a measurement of maximum $S_x$ and $S_y$ sensitivity, respectively. These measurements are averaged for each of the 19 scans, resulting in an experimentally determined dependence of the maximum $S_x$ and $S_y$ sensitivities on the angle between the filament and the QPD $y-$axis (Fig. \ref{Fibrils}d). We similarly compute the sensitivity values predicted by the bead model. The results are rescaled such that the largest measured $S_x$ sensitivity value was set to 1. As expected, the $S_x$ maximum sensitivity shows a maximum at a filament-QPD $y-$axis angle of $0^\circ$ and a minimum at an angle of $90^\circ$. Conversely, the $S_y$ maximum sensitivity shows a maximum at $90^\circ$ and a minimum at $0^\circ$. The maximum $S_y$ sensitivity at $90^\circ$ is approximately $30\%$ smaller than the maximum $S_x$ sensitivity at $0^\circ$. This same effect is seen in images of a single particle (Fig. \ref{SimVsRealResp}a) and is due to instrument-specific parameters.

The agreement between the model and the experimental data confirms that complex structures can be modeled as sums of point particles. Furthermore, for a uniform filament, the sensitivities depend not only on the filament's diameter but also on its orientation. With an understanding of these dependencies, we are now able to compare the thickness of filament-shaped objects regardless of their orientation in the $XY$-plane of the sample. By using one high quality template of a small bead, the instrument-specific angular-dependent response can be accounted for allowing for accurate modeling of the angle-dependent filament response using the ``bead model''. For networks that may contained out-of-plane tilted filaments, the out-of-plane angular dependence can be computed using a three dimensional scan of a single bead. In order to compare the diameter of two filaments positioned at different angles relative to the QPD axis, the following procedure is performed:

\begin{enumerate}
    \item A single small reference particle is scanned to be used as a reference point-particle (analogous to Fig. \ref{SimVsRealResp}a). Note, this reference can be used for many experiments as along as the alignment of the optical setup does not change.
    \item Using the point-particle reference, a ``bead model'' is generated to model the angular-dependent response accounting for the alignment of your optical setup (analogous to Fig. \ref{Fibrils}d).
    \item The height of both filaments is established (see suppl. sec. S5) and both filaments are scanned.
    \item The maximum sensitivity of both filaments is calculated. For filaments oriented closer to the QPD y-axis, $S_x$ maximum sensitivity is computed for each row of the scan. For filaments oriented closer to the QPD x-axis, $S_y$ maximum sensitivity is computed for each column of the scan. The orientation of the filament is computed using the locations of maximum sensitivity.
    \item Using the angular-dependent response of your setup (Fig. \ref{Fibrils}d), divide the maximum sensitivity measured for each filament by the correction factor for each filament's orientation. The resulting corrected maximum sensitivity, $\left(\partial S_x/\partial x\right)_{max,corr}$, is the sensitivity that would have been measured had that filament been oriented along the QPD y-axis.
\end{enumerate}.

By using this procedure BFP differential imaging becomes a quantitative method for comparing diameters of imaged filaments. In the next section, we demonstrate how the radius profile of a collagen fibril can be determined using this technique.

\section{Determining the profile of Collagen Fibril $\alpha-$Tips}

To demonstrate our ability to measure variations in the shape of filaments, we image the tapered $\alpha-$tips of collagen fibrils. Collagen fibrils are filaments $50\ -\ 500\ nm$ in diameter that provide mechanical rigidity to all connective tissue \citep{ECMrev_NatRevMolCellBio2014,ECMrev_AdvDrugDel2016}. Collagen fibrils self-assemble out of individual collagen proteins via an entropic process\citep{colHeirStruct_2011, colMimickHeir_2021}. Although the relative placement of neighboring molecules in fibrillar collagen has been known since the 1960s \citep{colFibStruct_1960, colDBandAFM_2022}, the exact subfibrillar structure \citep{subFibStruct_1974,subFibStruct_BASELT1993,subFibStruct_2011,subFibStruct_2013} and the mechanism of entropic growth \citep{alphaEnd_EBeam_1992,alphaEnd_rev1996,alphaTipModel_2019,colFibAssemb_HOLMES2018,colFibAssemb_2021} are active areas of research.

Type I collagen fibrils have been shown to possess two types of ends called $\alpha$ and $\beta$-tips. While $\beta$-tips have inconsistent profiles, $\alpha$-tips show a consistent linear increase in mass per unit length of approximately $0.2\ kDa/nm^2$ in electron microscopy studies \citep{alphaEnd_EBeam_1992, HolmesKadler_ColAlphaTip_JMolBio1998, alphaEnd_rev1996, colFibAssemb_2021} indicating a parabola-of-revolution tip shape. Fibrils $\alpha-$tips will show this parabola-of-revolution increase to their full fibril radius over the course of multiple $\mu m$. The shape of the $\alpha-$tip is used as an important clue for building models of molecular arrangements and understanding the mechanism of fibril growth \citep{alphaTipModel_2019}. Given their consistent tip shape, collagen $\alpha-$tips provide an excellent test platform for quantitative BFP microscopy of filamentous structures with changing mass per unit length.

\begin{figure}[htbp]
\centering\includegraphics[width=13cm]{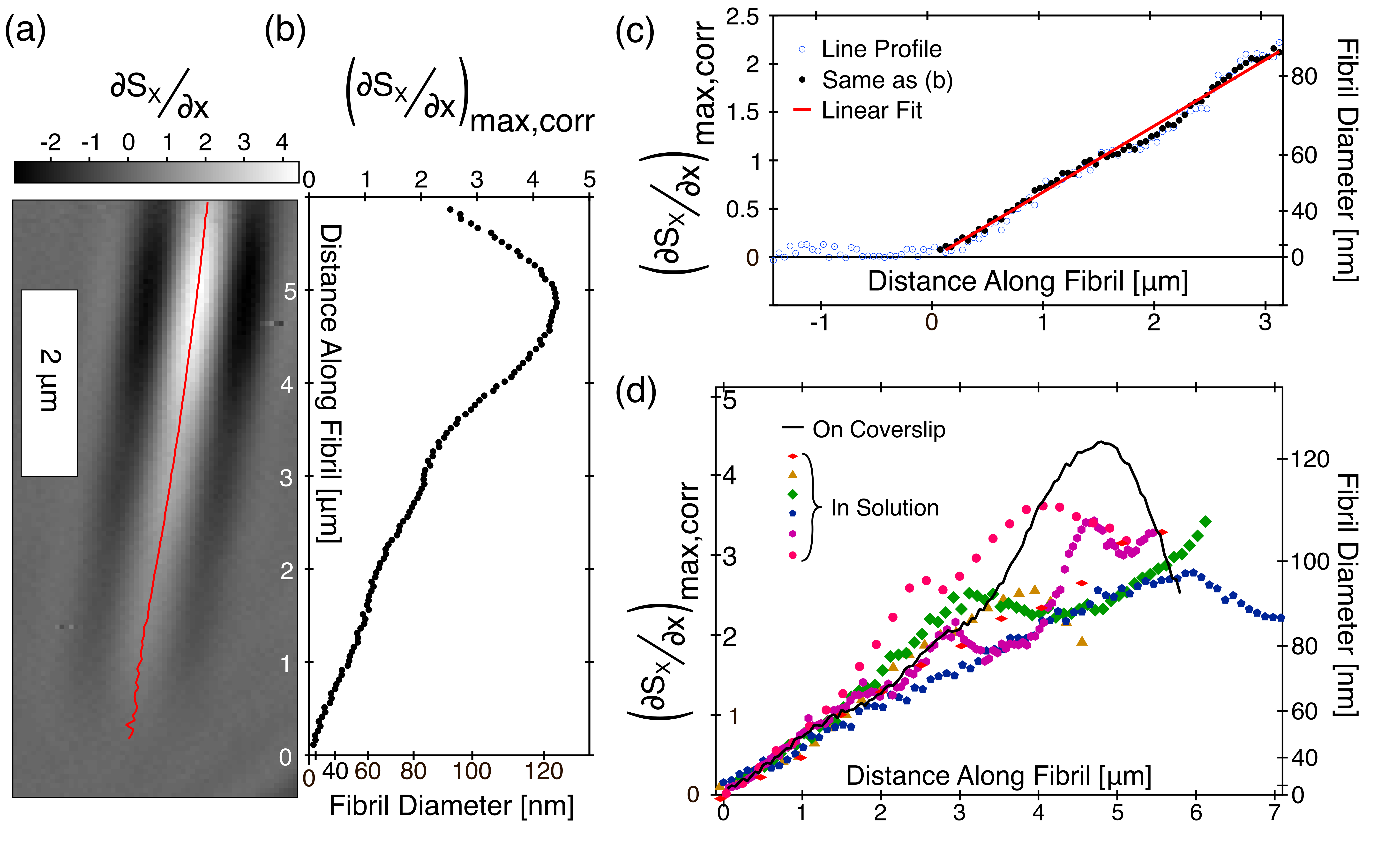}
\caption{BFP microscopy scans of collagen $\alpha-$tips. \textbf{(a)} $S_x$ sensitivity of a scan of a single collagen fibril on a coverslip surface. The location of maximum $S_x$ sensitivity coincides with the location of the fibril's axis and is marked by the red line. \textbf{(b)} The maximum $S_x$ sensitivity as a function of length along the fibril starting at the fibrils tip. The bottom axis shows the computed fibril diameter that corresponds to each maximum sensitivity measurement. \textbf{(c)} Magnified view of the maximum $S_x$ sensitivity (left axis) and corresponding fibril diameter (right axis) at the fibril tip. Maximum $S_x$ sensitivity computed using a parabolic fit  as shown in (b) (\scalebox{1.5}{$\bullet$}) is displayed along with the maximum $S_x$ sensitivity computed by taking a line profile along the center of the fibril (\textcolor{blue}{$\odot$}). A linear fit to the data is shown as a red line (\textcolor{red}{\textbf{\textemdash}}). \textbf{(d)} Maximum $S_x$ sensitivity measurements (left axis) and corresponding fibril diameter (right axis) for multiple collagen fibrils. Fibrils marked ``In Solution'' are attached to an electron microscopy carbon grid more than $20\ \mu m$ above the coverslip surface. The maximum $S_x$ sensitivity of these fibrils is plotted along with the maximum $S_x$ sensitivity of the coverslip-bound fibril (marked ``On Coverslip'') shown in parts (a), (b), and (c). In all parts of this figure, the $corr$ subscript to a maximum sensitivity indicates that the values have been corrected for filament orientation relative to the QPD $y-$axis. All sensitivity values have been rescaled such that the $(\partial S_x /\partial x)_{max}=1$ for a $110\ nm$ diameter polystyrene bead. }
\label{colAlphaEnd}
\end{figure}

To investigate the profile and mass distribution of $\alpha-$tips, we scan the tip of a reconstituted collagen fibril attached to a coverslip surface. Fig. \ref{colAlphaEnd} (a) shows the $S_x$ sensitivity of the scan computed by taking the derivative $\partial S_x / \partial x$ at each location of the scan. The sensitivities are scaled relative to the maximum $S_x$ sensitivity of a $110\ nm$ bead. We find the maximum sensitivity at sub-pixel resolution of each row by fitting a parabola to a $0.4\ \mu m$ wide region around the peak. The locations of the maximum $S_x$ sensitivities, corresponding to the location of the fibril's axis, are plotted as a red line. Fig. \ref{colAlphaEnd}b shows the maximum $S_x$ sensitivity along the fibril corrected for the filament's orientation relative to the QPD $y-$axis using the calibration curve in Fig. \ref{Fibrils}d, $\left(\partial S_x/\partial x\right)_{max,corr}$. Here, the subscript `$corr$' indicates that the maximum sensitivity has been corrected for rotation with respect to the QPD $y-$axis using the model in Fig. \ref{Fibrils}d. I.e. the presented value represents the maximum $S_x$ sensitivity that would be measured for an identical filament that was aligned with the QPD $y-$axis. To compute the corresponding fibril diameter profile, we use the ``bead model'' to construct a fibril of diameter $d_{bm}$ aligned with the QPD $y-$axis and compute the corresponding maximum $S_x$ sensitivity, $\left(\partial S_x^{bm}/\partial x\right)_{max}$. To compare the ``bead-model'' to the collagen fibril, we have to account for differences in index of refraction between collagen and the reference bead used as a template for the ``bead model''. In order to compute the diameter of the collagen fibril, $d_{cf}$, we recall that for a cylindrical object, the sensitivity is proportional to the square of the diameter. Therefore:

\begin{align}
    \dfrac{d_{cf}}{d_{bm}} &= \dfrac{\sqrt{\left(\partial S_x/\partial x\right)_{max,corr}}}{\sqrt{\left(\partial S_x^{bm}/\partial x\right)_{max}\times n_{corr}}} \label{eqn:colDiam}\\
\end{align}

For this paper, we constructed a ``bead-model'' using $110\ nm$ diameter polystyrene beads. Therefore, we use the simulated response of a polystyrene cylinder with diameter $d_{bm}=110\ nm$ as a reference. The index of refraction of polystyrene ($n_{bm} = 1.57$), the refractive index of collagen ($n_{cf}$ in the range of $1.411-1.418$\citep{col_RefrInd_Leonard1997}), and the refractive index of the surrounding solution ($n_s = 1.333$) are used in our calculation. Note that the index of refraction of collagen presented here differs significantly from the index of refraction of most proteins. The value used gives the refractive index of hydrated collagen and includes the dry collagen fibril and the hydration layer present in the collagen under physiological ion concentration. By using this value for the index of refraction, we effectively account for the density of hydrated collagen and are able to compute the diameter of the fibril. Using these values, we simplify the above equation to:

\begin{align}
    d_{cf}&=A\times \sqrt{\left(\partial S_x/\partial x\right)_{max,corr}}
    \label{eqn:OurColD}\\
    & \qquad \mathrm{where} \quad A=58.6\ nm \notag
\end{align}

\noindent Note that the value of the prefactor $A$ depends on instrument parameters and needs to be determined independently by scanning a reference object. Using Eqn. \ref{eqn:OurColD}, we can now calculate the fibril diameter of the imaged collagen fibril (Fig. \ref{colAlphaEnd}b, bottom axis). Note that the scales for the diameter axes in b, c, and d are not linear.

The lower $3\  \mu m$ of the fibril show a constant increase in the maximum $S_x$ sensitivity indicating a linear increase in the cross-sectional area of the fibril. This is consistent with a linear growth in the mass per unit length observed in electron microscopy \citep{alphaEnd_EBeam_1992, alphaEnd_rev1996, HolmesKadler_ColAlphaTip_JMolBio1998}. A magnified view of the lower $3\ \mu m$ of the fibril is shown in Fig. \ref{colAlphaEnd}c. Toward the tip end, the linear decrease in fibril cross-sectional area continues until the signal vanishes into the background. Using the standard deviation of the background as the limiting factor, we can resolve the collagen fibril's $\alpha-$end until it shrinks to a diameter of $13\ nm$. As this BFP microscopy methods is not limited to the surface, we repeat the measurements of $\alpha-$tip diameter profiles with collagen fibrils fixed to a electron microscopy holey carbon grid at a height of $20\ um - 80\ um$ above the surface of the coverslip. The maximum sensitivities for the suspended fibrils are shown in Fig. \ref{colAlphaEnd}d. To select only for $\alpha-$tips, collagen fibrils tips that showed an irregular and abrupt end were excluded from the dataset. All collagen fibril $\alpha-$tips display a consistent shape over the first $\approx 2\ \mu m$ before some filaments' maximum sensitivity profiles begin to diverge due to difference in their absolute fibril diameters and artifacts from the filament preparation protocols. Being able to resolve the shape of the $\alpha-$tips with similar precision regardless of their position in the sample underlines the ability of our analysis method to extract structural information such as mass distributions anywhere in the sample.

\section{Background Subtraction and Linear Density of Microtubules}

In many assays that analyze surface bound nanostructures such as thin filaments, imaging is limited by co-adsorption of protein aggregates or glass imperfections rather than instrument noise leading to low signal-to-background ratios (SBR). Due to the angle-dependent responses created by the use of a QPD, a unique method of background subtraction may be implemented. While a point-like object such as a small particle or protein aggregate is visible in both the $S_x$ and $S_y$ channels (Fig. \ref{SimVsRealResp}), a uniform filament oriented along the QPD $x-$axis remains invisible in the $S_y$ channel (Fig. \ref{Fibrils}). Therefore, the $S_y$ signal can be used to identify the location and size of point-like objects. Since, for the microscopy method presented here, signals are additive for small particles, the contribution to the $S_x$ signal originating from point-like objects can be subtracted strongly reducing the signal-to-background ratio of the resulting image.

To demonstrate the advantage of this unique background reduction method, we use microtubules. As the stiffest cytoskeletal filament, microtubules provide structure, organelle organization, facilitate active transport, and aid in cell division in eukaryotic cells \citep{MolBioOfCell_2002,transAlongMTrev_Welte2004, MitochondTransp_MacAskillKittler_TrendsCellBio2010,cytoSkelRev_Fletcher2010, MechOfMicrOrg_Akhmanova_NatRevMolCellBio2022}. Due to their many vital roles, microtubule mechanical properties and structure are an active area of research \citep{MTPlusEnds_JHow2003, MTPlusEnd_Igaev2022, AsymKinetochore_ChipAsbury_JCellBio2024}. Microtubules are hollow cylindrical structures that most commonly have an outer and inner diameter of $\approx 25nm$ and $\approx 15nm$ respectively \citep{MTstruct_Pampaloni2008}. Microtubules consist of tubulin dimers assembled into individual protofilaments. These protofilaments in turn assemble into a cylindrical structure. While 13 protofilament microtubules are most commonly found in the mammalian cytoskeleton, the number of protofilaments can vary depending on the preparation method when assembling the microtubules in vitro \citep{MTBestiary_Brouhard_MolBioCell2017}. In vitro microtubule polymerization assays, like the one used here, produce microtubules ranging from 9 to 16 protofilaments with 12, 13, and 14 protofilament microtubules being most common \citep{MT_PFnum_JCellBio1978,MT_LatticePFnum_Wade_JCellBio1992}. The number of protofilaments not only changes the diameter, but also changes the structure and mechanical behavior of the microtubule \citep{Supertwist_Flyvbjerg_BioOfCell2007, WLB_EFrey_PhysRevE2010, Supertwist_Schaffer2018}. Due to their thin size, fluorescence microscopy is often used for visualizing microtubules. However, quantifying microtubule linear density with fluorescent labels is difficult due to photo-bleaching and uneven label densities. Furthermore, fluorescent markers may affect microtubule structure and function \citep{FluorRuinsMT_Traffic2015} underlining the need for label-free imaging. As thin, rigid filaments of uniform thickness, microtubules provide an excellent platform for testing our unique background subtraction technique.

\begin{figure}[htbp]
\centering\includegraphics[width=10cm]{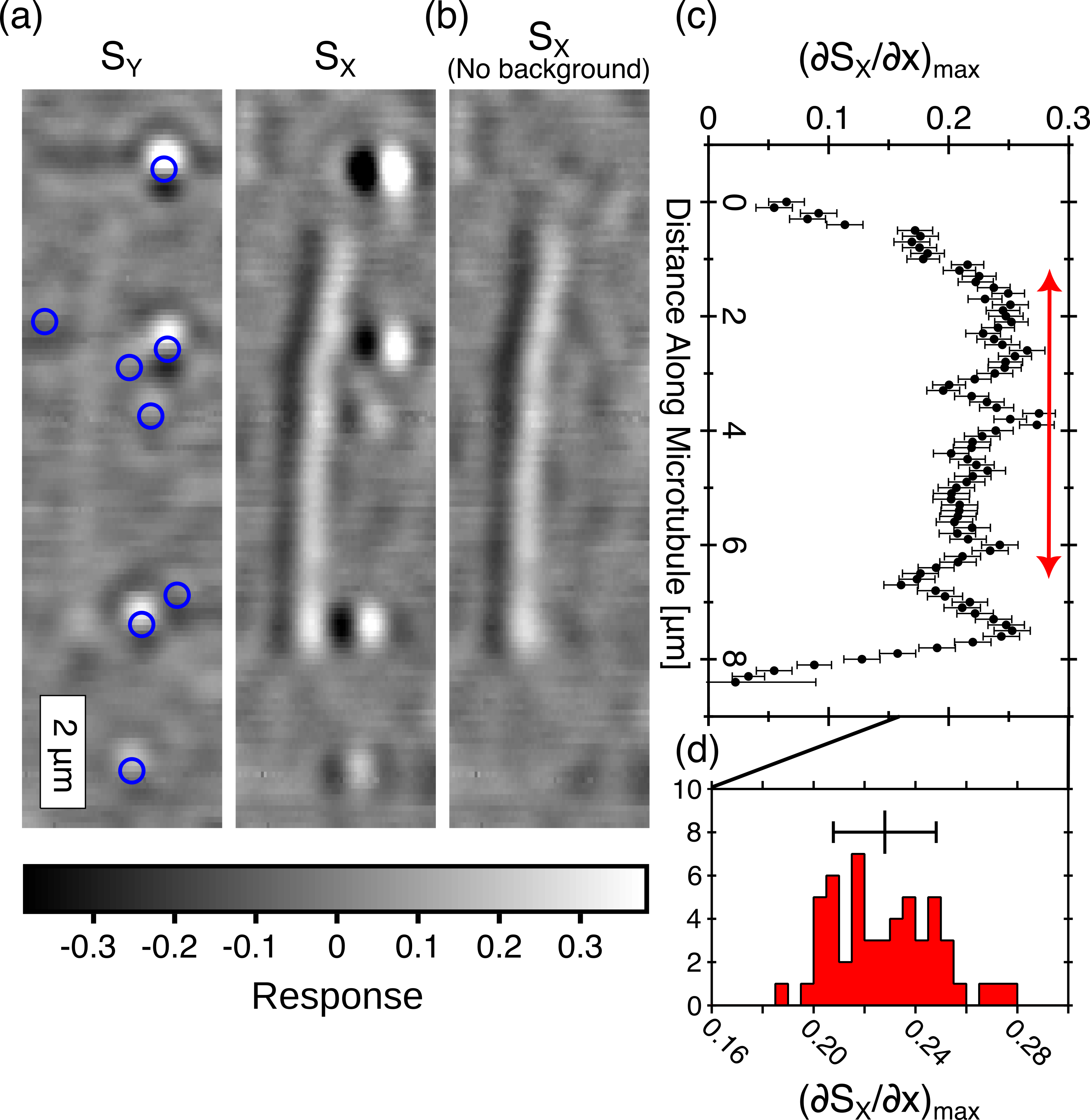}
\caption{Background subtraction method for point-like objects. \textbf{(a)} $S_y$ and $S_x$ responses of a microtubule along with protein aggregates bound to a PLL coated coverslip. Blue circles (\textcolor{blue}{\scalebox{1.5}{$\circ$}}) indicate the positions of protein aggregates to be removed. \textbf{(b)} $S_x$ response after the removal of the protein aggregate background. \textbf{(c)} Maximum sensitivity along the microtubule computed after the removal of background protein aggregates. The red arrow indicates the region of uniform maximal $S_x$ sensitivity for the microtubule. \textbf{(d)} Histogram of maximal $S_x$ sensitivity values in the region of uniform $S_x$ sensitivity. The mean and standard deviation are shown above the histogram.}
\label{NoiseSubMT}
\end{figure}

A sample with coverslip-bound microtubules is created as described in Materials and Methods. Fig. \ref{NoiseSubMT}a shows representative $S_y$ and $S_x$ images of a microtubule with protein aggregates surrounding the filament. Note that the microtubule shows maximum sensitivity in the $S_x$ signal and is almost invisible in the $S_y$ signal due to the microtubule's alignment with the QPD $y-$axis. However, the protein aggregates are visible in both the $S_x$ and $S_y$ channels, because, being point-like objects, their response is identical to the response of a small particle.  If $n$ protein aggregates are visible in the $S_y$ response, a weighted sum of $n$ translated point-like object $S_y$ templates (Fig. \ref{SimVsRealResp}:middle ) is fit to the $S_y$ response analogously to the fit described in Eqn. \ref{eqn:multiBeadFit}. Once the locations and relative sizes are identified by the fit, $n$ point-like object $S_x$ templates with the same locations and relative sizes can be subtracted from the $S_x$ response (Fig. \ref{NoiseSubMT}b ). While the point-like protein aggregates dominate the contrast in the raw image, the microtubule becomes the prominent object in the background-subtracted image. The remaining background visible in Fig. \ref{NoiseSubMT}b is a result of an uneven surface coating and smaller aggregates whose responses are too noisy to be fit accurately with a point-particle template.

Microtubules are expected to have a constant radius along multi-micron length scales \citep{MT_LatticePFnum_Wade_JCellBio1992}. Microtubules may have uneven ends, in particular curling protofilaments at the microtubule's plus end \citep{MT_curlTips_1991} which could affect the uniformity of the signal. To exclude effects caused by the microtubule ends and the width of the beam, we restricted the analysis to the region at $1.2\mu m$ away from the first non-zero maximum $S_x$ sensitivity measurement (see red double-sided arrow in Fig. \ref{NoiseSubMT}c ). The histogram of the maximum $S_x$ sensitivities is shown in Fig. \ref{NoiseSubMT}d. Using the maximum $S_x$ sensitivity of the background-subtracted $S_x$ response along the length of the microtubule (Fig. \ref{NoiseSubMT}c ), we can compute the linear density of the measured microtubule and quantify the improvement in image quality by computing the SBR. In order to compute the linear density of the microtubule, $\mu_{mt}$, we use a ``bead model'' to construct the response of a polystyrene cylinder aligned with the QPD $y-$axis of fixed linear density $\mu_{bm}$. Analogously to equation \ref{eqn:colDiam}, sensitivity is proportional to linear density which allows us to write the following relationship:

\begin{align}
    \dfrac{\mu_{mt}}{\mu_{bm}} = &\dfrac{\rho_{mt}}{\rho_{rb}}\times\dfrac{\left(\partial S_x^{mt}/\partial x\right)_{max,corr}}{\left(\partial S_x^{bm}/\partial x\right)_{max}\times n_{corr}} \label{eqn:mtLinDens}
\end{align}

Using an index of refraction for dry globular proteins for the index of refraction of the microtubule, $n_{mt} = 1.587$ \citep{iScatProtMass_Science2018}, the index of refraction of polystyrene, $n_{bm} = 1.57$, the index of refraction of the microtubule buffer, $n_s=1.335$, we measure this microtubule's linear density to be $\mu_{mt}=169\ MDa/\mu m$. Given the molecular weight of a tubulin dimer, $110\ kDa$, and the length of a tubulin dimer, $8\ nm$, we can approximate the linear density of a protofilament to be $13.7\ MDa/\mu m$. Therefore, 12 protofilament microtubule would have a linear density of $164\ MDa/\mu$, most closely matching our measurement.

We also use the sensitivity profile to estimate the improved SBR. The magnitude of the signal is given by the average maximum sensitivity of the microtubule. As the microtubule is superimposed on a point-particle background, in order to estimate the magnitude of the background, we compute the standard deviation of the sensitivity in a region away from the microtubule. We now estimate the SBR by computing the magnitude of the microtubule signal divided by the magnitude of the point-like particle background. Computed in this way, we find an SBR improvement from $3.1$ to $11.8$ for the microtubule in Fig. \ref{NoiseSubMT}. Alternatively, the SBR can be estimated by computing the ratio of the average maximum $S_x$ sensitivity to the standard deviation in maximum $S_x$ sensitivity of the microtubule. The SBR calculated in this way is $11.3$ (Fig. \ref{NoiseSubMT}d. The reproducibility of the SBR using both methods supports the assumption that the microtubule has a uniform thickness. Applying the first method for computing the SBR to other microtubule scans, we consistently achieve a SBR in the range of $10-12$ after performing the background subtraction despite initial SBRs in the range of $3-6$. Despite starting with a sample containing many protein aggregates, we are able to achieve a SBR comparable to the best label-free microscopy methods \citep{IntReflMicrSBR_ESchaf_JHow_2018, iScat_SBR_Kukura2016}. With this, determining the molecular architecture, such as the number of protofilaments, of microtubules is within reach.

\section{Discussion}

Here we introduced a BFP differential scanning microscopy technique for imaging sub-diffraction sized biological filaments and measuring their radii and linear densities. This technique extends the current uses of DPC microscopy through a data analysis technique based on highly precise optical tweezer BFP particle position tracking. Therefore, this technique can be implemented in current optical tweezer systems already equipped with BFP position detection with no modifications to the experimental setup with the exception of an extra ND filter in the case where trapping forces need to be minimized. This adds a powerful new imaging capability to optical tweezers.

We demonstrate mathematically and experimentally that the resulting microscopy image of small objects can be constructed as the sum of images of single Rayleigh scatterers. By using a single small reference bead, the measurement of mass distributions of more complex images already accounts for instrument specific parameters. Using this technique, we showed that our system is capable of measuring mass on the order of $1.1\ MDa$ for globular proteins and measuring collagen fibril $\alpha-$tip diameter profiles down to a diameter of $13\ nm$. We further demonstrated a unique background subtraction technique made possible by the simultaneous collection of two images. By using microtubules as a benchmark, we measured microtubule linear mass density with an SBR of $10-12$.

\subsection{Further Improvements}

In this work, a laser wavelength of $1064\ nm$ was used. This wavelength is common for optical tweezer applications to minimize photodamage to biological materials \citep{KNeumanSBlock_Photodamage_BiophysJ1999}. However, the use of shorter wavelengths of light can improve both the spatial resolution and the mass resolution. The spatial resolution is governed by the beam width at the focus. As this is a diffraction limited system, the beam width is approximated by $w_0 \approx \lambda/(2\times n_s)$ where $\lambda$ is the wavelength of the light and $n_s$ is the index of refraction of the sample medium. As the wavelength decreases, the spatial resolution of the imaging method is improved. Furthermore, the intensity of the measured signal is proportional to the scattered electric field, $E_{scat}$ (eqn. \ref{eqn:ReEEpReEE}. For a Rayleigh scatterer, $E_{scat}\propto1/\lambda^2$ (see suppl. sec. S3). Therefore, decreasing the wavelength would give a quadratic increase in the measured signal resulting in better mass resolution. Considering the low light intensity used by our system ($0.13\ mW$ compared to the $100\ mW$ commonly used for optical tweezer setups \citep{KNeumanSBlock_Photodamage_BiophysJ1999}), using a stable shorter-wavelength laser could significantly improve spatial and mass resolution.

\subsection{Comparison to Other Label-Free Methods}

Single, label-free imaging of collagen fibrils and microtubules has been accomplished with other techniques. Differential interference contrast (DIC) is a commonly used label-free method. DIC relies on the interference between light that passes through two nearby points in the sample. Microtubules measured with DIC show an SBR of 10.4 \citep{IntReflMicrSBR_ESchaf_JHow_2018}. DIC is similarly able to measure far away from the coverslip. Also, like the QPD based method presented here, DIC shows orientation dependence. The Wollaston prism shear axis establishes the offset between the two light paths interfered on the detector. For filament-like objects, this results in a maximum contrast orientation. Filaments orthogonal to this orientation become invisible. This orientation dependence in DIC can be negated by either taking multiple images at different angles, or by changing the Wollaston prism shear axis \citep{ShribakInou_OrientFreeDIC_ApplOpt2006}. Both implementations require multiple images and changes to the data taking protocol and microscope setup. While the QPD based BFP measurement presented here also shows orientation dependence, no filament orientation is invisible and the orientation dependence can be easily accounted for using the method described in section \ref{sec:ImFil}.

Second harmonic generation (SHG) microscopy has been used to measure fibril diameter to within $30\ nm$ \citep{Bancel_SHGColRad_NatCom2014}. SGH has the benefit of providing extra information about the internal arrangement and structure of collagen bonds. However, the study was performed using dried collagen and orders of magnitude higher laser power required increased the risk of photodamage. In comparison, the method presented here allows for identifying fibril diameters within $13\ nm$ for fibrils under physiological conditions while using very lower laser power.

Interferometric reflection microscopy (IRM) and interferometric scattering microscopy (iSCAT) both work through interference between the scattered light and the light reflected by the coverslip-solution interface. By using the reflection of the coverslip, the background intensity is decreased significantly improving the signal-to-noise ratio \citep{iScat_NanoLet_2019}. Both techniques have been used to image microtubules \citep{IntReflMicrSBR_ESchaf_JHow_2018, iScat_SBR_Kukura2016}. IRM achieves a SBR for single microtubules of $6.8\pm0.8$ for single image with further improvement to the SBR from averaging and post-processing \citep{IntReflMicrSBR_ESchaf_JHow_2018}. IRM has the benefit of being a relatively simple, wide-field imaging system. However, the use of a camera limits IRM temporal resolution while the laser scanning method presented here allows for collecting low-noise single-pixel intensity data at a MHz sampling rate. iSCAT is also a laser scanning technique and allows for extremely high sampling rate measurements at one pixel position. With better control of the illumination, iSCAT has been used to measure single microtubules with an SBR of $26$ \citep{iScat_SBR_Kukura2016}. When applied to single proteins arriving at the surface, with the use of frame subtraction and other machine learning driven post processing techniques, iSCAT can achieve a mass sensitivity of $10\ kDa$ \citep{VSandoghdar_10kDaiScat_NatMeth2023}. However, this high mass sensitivity relies on frame subtraction and is only achieved for single proteins at the moment they arrive at the coverslip surface. This level of accuracy has not been demonstrated for permanent structures in the sample. Also, the reliance on the reflection from the coverslip surface limits both techniques to measure near the coverslip. In contrast, using BFP detection allows for measurements deep within the sample making it possible to study complicated structures such as networks and avoid surface effects.

\subsection{Outlook}

As the microscopy method presented in this work allows for measurements deep within the sample volume, this method can be used to image three dimensional structures such as filament networks. Paired with high-bandwidth, low-noise detection, network dynamics can be studied in parallel with imaging the filament positions and thickness profiles. For example, the effects of stress on tension within artificial ECMs \citep{LissekBiorX2018, Doyle_DeformECMbyCell_DevCell2021} can be studied to explain tissue properties and forces relevant to cell mechanotransduction \citep{MechanotransRev_NatRevMolCellBio2023}. This technique can also be used to study mechanical properties of single microtubules and collagen fibrils. With the rise of studies using optical tweezers, we expect this microscopy method to become a powerful tool for studying the connection between structure and mechanics in microscopic filament systems.

\section{Experimental Methods}

\subsection{Coverslip Cleaning Procedure}
Coverslips (\#1.5 Round 15mm Thermo Fisher Scientific or Paul Marienfeld GmbH \& Co.KG) are placed on a coverslip cleaning rack and the rack is placed in a staining dish. The coverslips go through three washing cycles. For each washing cycle, the staining dish is filled with Hellmanex-III (9-307-011-4-507, Hellma, USA) solution diluted to $2\%$ in pure deionized water from a Milli-Q water system and sonicated for 15 minutes. The staining dish is then rinsed with deionized water from the Milli-Q water system for 5 minutes to remove the Hellmanex solution without allowing any part of the coverslips above the surface of the washing solution. The coverslips are then sonicated in Milli-Q water for 15 minutes. After another 5 minute rinse with deionized water, the coverslips are removed from the water and placed into fresh $2\%$ Hellmanex solution to start a new washing cycle. After a total of three washing cycles, the coverslips are dried on the rack using nitrogen gas and stored in a nitrogen gas-filled staining dish sealed with Parafilm (PM996, Bremis).

\subsection{Collagen Polymerization}
The collagen polymerization procedure was based on a procedure used by \textit{Mickel et al.} \citep{ColProtocol_Mickel2008}. Type-I rat tail tendon (354236, Corning) and type-I bovine dermis collagen (354231, Corning) were mixed at a relative concentration of 1:2. The collagen concentration was diluted to a total concentration of $0.8\ mg$ collagen protein per ml of a 1:1 solution of 10x DMEM and $0.27\ M$ \ce{NaHCO3}. The collagen, and 10x DMEM and \ce{NaHCO3} solution is stored at $4^\circ C$ and mixed on ice. The resulting solution has a pH of 10. $100\ \mu L$ of the collagen solution was then placed into a $1.5\ mL$ centrifuge tube and placed into an incubator at high humidity, $37^\circ C$ and $5\%$ \ce{CO2} by volume for an hour. After an hour, a loose collagen network is formed in the centrifuge tube. $1\ mL$ of 1X PBS (D8537, Sigma Aldrich) is added to the centrifuge tube and the network is broken up via vigorous pipetting using a $1\ mL$ pipette tip with a cutoff end. The end is cut off in order to avoid excessive shear forces in the fluid. After the collagen network has been broken up, it is ready to be further diluted in 1X PBS and added to the microscopy sample.

\subsection{Microtubule Polymerization}
Microtubules are polymerized from a 1:5 ratio of Rhodamine-labeled to unlabeled porcine-brain tubulin (respectively TL590 and T240, Cytoskeleton). The unlabeled tubulin is stored in $10\ \mu L$ aliquots with $10\ mg/mL$ tubulin, $4\ mM$ \ce{MgCl2}, $1\ mM$ GTP (G5884, Sigma-Aldrich) in Brinkley Renaturing Buffer 80 (BRB80) at $-80^\circ\  C$. BRB80 consists of $80\ mM$ PIPES (P6757,
Sigma-Aldrich), $1\ mM$ EGTA (E4378, Sigma-Aldrich), $1\ mM$ \ce{MgCl2}, pH 6.8). Note, the total concentration of \ce{MgCl2} is $5\ mM$ as this increases tubulin's affinity for GTP \citep{MT_MgCl2_1990}. The Rhodamine-labeled tubulin is stored at $4^\circ C$ as a lyophilized powder. The Rhodamine-labeled tubulin is dissolved is $15\mu L$ of $1\ mM$ GTP and $5\ mM$ \ce{MgCl2} in BRB80 and added to the thawed aliquot of unlabeled tubulin. During this process the tubulin is always kept on ice. Then the tubulin solution is centrifuged at $90,000\ RPM$ for $5\ min$ while remaining at $4^\circ C$ using a TLA-100 rotor in a temperature-controlled ultracentrifuge (Beckman TL100, Beckman Coulter). As a result, misfolded tubulin proteins aggregate into a pellet at the bottom of the centrifuge tube. The supernatant solution was incubated in a $37^\circ C$ water bath for $15$-$20\ min$. A second $5\ min$ centrifuge spin at $40,000\ RPM$ and $37^\circ\ C$ is used to aggregate the polymerized microtubules into a pellet. A pipette is used to gently wash the microtubule pellet 3-5 times using $50\ \mu L$ of BRB80 with $20\ \mu M$ Taxol (T1912, Sigma-Aldrich). Then, using a pipette tip with a cut end to avoid excessive shear forces, the microtubule pellet is resuspended in $50\ \mu L$ of BRB80 and $20\ \mu M$ Taxol. Before the start of an experiment, the microtubule solution is further diluted to the needed concentration in a Pyranose Oxidase and Catalase oxygen scavenging system of $3\ Units/mL$ of Pyranose Oxidase (P4234, Sigma-Aldrich), $90\ Units/mL$ of Catalase (C40, Sigma-Aldrich), and $44mM$ Glucose in BRB80 with $20\mu M$ Taxol.

\subsection{PLL Coverslip and Carbon Grid Surface Coating}
In order to attach microtubules, or collagen to the coverslip, the bottom coverslip or electron microscopy carbon grid (Quantifoil S 7/2 Holey Carbon Grids on Gold Foil) is coated in poly-L-lysine (PLL). To ensure a clean hydrophilic surface, a cleaned coverslip or carbon grid is placed in a ceramic coverslip rack and plasma-cleaned for 1 minute in a nitrogen plasma using a tabletop plasma cleaner (PDC-001, Harrick Plasma). A PLL solution is prepared by diluting $0.1\%$ w/v PLL (P8920, Sigma Aldrich) by 200 times in ethanol. The plasma-cleaned coverslip or carbon grid is placed in the PLL solution for 15 minutes, then allowed to slowly dry in air before being attached to the sample chamber.

\subsection{Sample Assembly}
The sample holders are either glass or titanium and have a cylindrical $1\ mm$ thick, $0.5\ in$ ($\approx 13\ mm$) diameter opening that contains the aqueous sample. The sample chamber is heated just enough to melt dental wax. Dots of liquid dental wax are placed around the cylindrical sample holder on one side and a plasma-cleaned coverslip is placed on top allowing the wax to spread between the sample holder and the coverslip via capillary action. When collagen or microtubules are attached directly to the coverslip, this coverslip is coated in PLL. The sample holder is quickly cooled to solidify the dental wax and fix the coverslip in place. If suspending collagen using a carbon grid, the PLL coated carbon grid is attached to the center of the bottom coverslip with two small dots of vacuum grease on diametrically opposed sides of the carbon grid.  Approximately $ 150\ \mu L$ of the microtubule or collagen solution is washed onto the sample. In the case of microtubules, they are allowed to settle onto the PLL coated surface for $30\ min$. In the case of collagen, the solution is pipetted in and out a couple of times while ensuring that the coverslip surface remains fully submerged to increase the collagen fibril concentration on the surface. The remaining solution is rinsed three to four times with $100\ \mu L$ of BRB80 and Taxol solution or PBS for microtubules or collagen respectively. Finally, dots of vacuum grease are placed around the remaining opening and a clean but not plasma-treated coverslip is attached. A Kimwipe (\#34120, Kimtech) is used to absorb excess liquid that has been squeezed out during the closing of the sample chamber. The sample is then placed in the optical tweezer setup for imaging.

\subsection{Fluorescent beads}
All carboxylated fluorescent beads were yellow-green FluoSpheres (F8888, Invetrogen). 

\nocite{Lasers_Svelto}
\nocite{FieldGuideToLasers_Paschotta}
\nocite{dielectricDispersion}
\nocite{gradForce_1996}
\nocite{RaylCross_AmJPhys2002}

\section{Data availability}
Data will be made available on request.

%%       \citet{<label>} ==> Lamport (1994)
%%       \citep{<label>} ==> (Lamport, 1994)
%%
%Example citation, See \citet{lamport94}.

\bibliographystyle{apalike}
%elsarticle-harv
\bibliography{theBib_OptTweezMicr}
%\printbibliography



%\begin{thebibliography}{00}


%\bibitem[Lamport(1994)]{lamport94}
%  \textit{\LaTeX: a document preparation system},
%  Addison Wesley, Massachusetts,

%\end{thebibliography}
\end{document}

\begin{frontmatter}



\title{Supplemental Materials: Back-Focal-Plane Imaging and Linear Density Measurements Of Sub-Diffraction Sized Biological Filaments and Particles} %% Article title


\end{frontmatter}




\section{Optical Path}
A detailed view of the optical path is presented in Fig. S1\iffalse\ref{OptPath1}\fi. A list of relevant microscope components is presented in Table S1. The $1064\ nm$ laser, after passing through a Faraday isolator and a beam expander, enters the diagram in Fig. S1 \iffalse\ref{OptPath1}\fi on the bottom indicated by the red arrow. A neutral density filter reduces the light intensity enough to avoid trapping (see calculation in Section S3). After the neutral density filter, a $45^\circ$ mirror directs the laser light into the objective lens which focuses the light in the sample. The condenser lens with an oil immersion attachment collects the scattered and unscattered laser light. Another $45^\circ$ mirror directs the light into a pair of convergent lenses that allow for further placement of neutral density filters. The resulting interference pattern between the scattered and unscattered light is projected onto the quadrant photodiode. During data collection, the piezoelectric three-axis stage is scanned through a series of positions. The stage stops at each position for at least $0.1$ seconds and the quadrant photodiode signals are recorded.

\begin{figure}[htbp]
\centering\includegraphics[width=10cm]{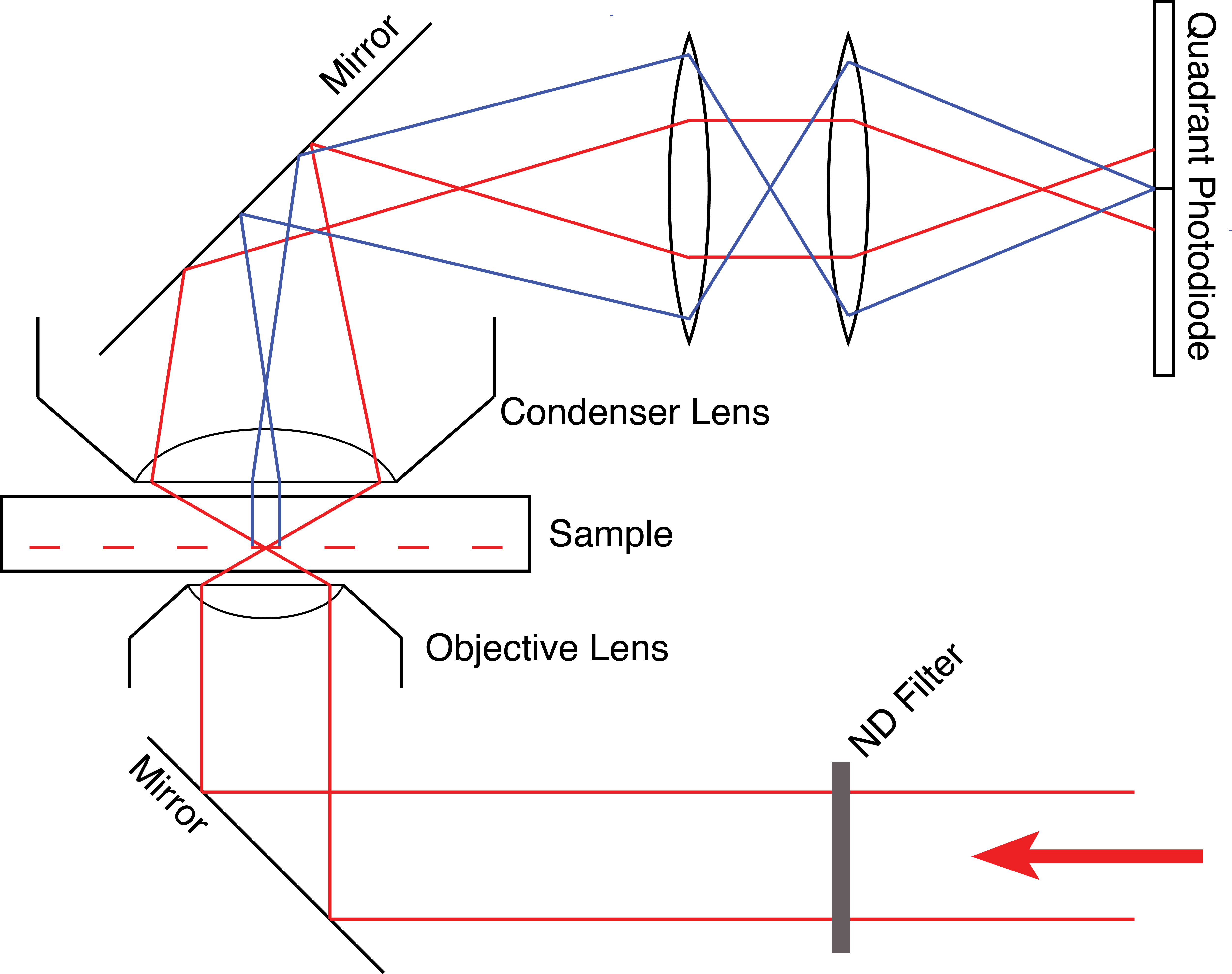}
\caption{Optical path for BFP differential microscopy used in this study. The red lines show the path of the laser light while the blue lines show the locations of planes conjugate with the back focal plane of the condenser lens.}
\label{OptPath1}
\end{figure}

A simplified view is helpful in computing the optical interference pattern seen by the QPD (Fig. S2\iffalse\ref{OptPath2}\fi ). After passing through the objective lens, the laser light enters from the left side of the image. Some of the light is scattered by the particle shown as a green circle. As the particle is located very close to the laser focus, the unscattered light ray and the light ray scattered by the particle (shown in red and green) that hit a single point on the condenser lens' principal surface will be traveling in the same direction. Light rays hitting a single point on the condenser lens' principal surface and traveling in the same direction will be mapped to a single point on the QPD, since the QPD is located in the back focal plane of the condenser. Therefore, it is sufficient to compute the interference pattern on the principal surface of the condenser lens (\CiteGittes).

\begin{figure}[htbp]
\centering\includegraphics[width=8cm]{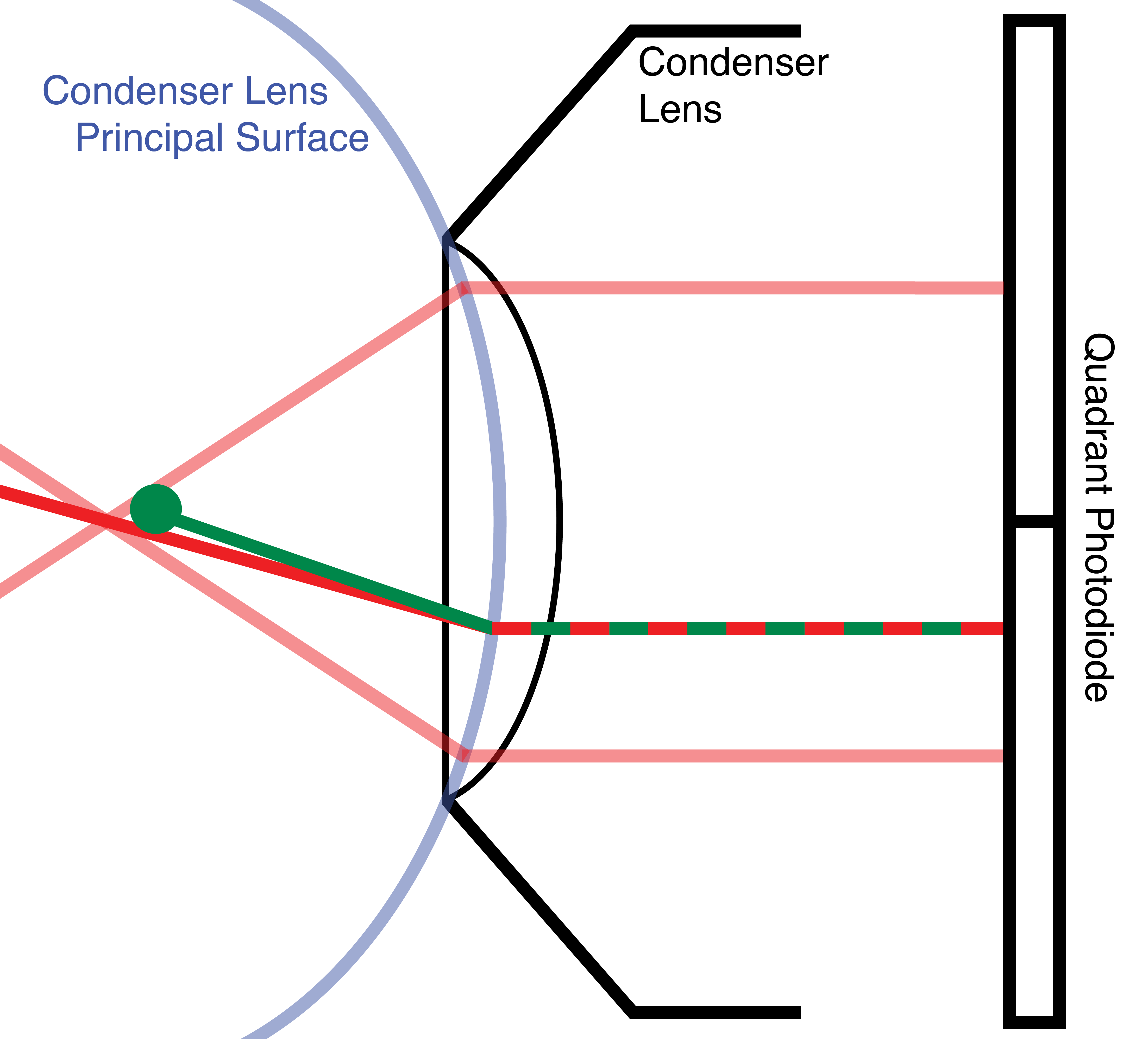}
\caption{Simplified optical path for calculating the interference pattern and QPD response. An example of a unscattered and scattered light ray interfering at the QPD is shown in red and green, respectively.}
\label{OptPath2}
\end{figure}

\begin{table}[h!]
    \centering
    \begin{tabular}{|c|c|}
    \hline
        \textbf{Component} & \textbf{Specifications} \\
    \hline
    \hline
        \multirow{2}{*}{Infrared Laser} & $500\ mW$, $1064\ nm$ Nd:YAG \\
          & Mephisto, Coherent, USA \\
    \hline
        \multirow{2}{*}{Faraday Isolator} & Electro-Optics Technology Inc.\\
         & $1030-1080\ nm$, $2\ mm$ Aperture \\
    \hline
        Beam Expander & \\
    \hline
        ND Filter & $50\ \%$ attenuation  \\
    \hline
        ND Filter & $95\ \%$ attenuation \\
    \hline
        \multirow{2}{*}{Shutter} & Unibilitz Vincent Associates \\
         & Electronic Shutter\\
    \hline
        Dichroic Mirror & \\
    \hline
        $60x$ $1.2NA$ \ce{H2O} Immersion Objective Lens & UPLSAPO60XW, Olympus, Japan \\
    \hline
        Water Immersion Oil & Immersol W 2010, Zeiss, Germany \\
    \hline
        2 Axis Manual Stage & Mad City Labs, USA \\
    \hline
        3 Axis Piezo Stage & PDQ375, Mad City Labs, USA \\
    \hline
        \multirow{2}{*}{Bottom Coverslip} & 152222, Epredia, Germany or \\
          & 15CIR-1.5, ThermoFisher, USA \\
    \hline
        Interior of Sample &  \\
    \hline
        Top Coverslip &  Same As Bottom Coverslip \\
    \hline
        Immersion Oil & Immersol 518F, Zeiss, Germany \\
    \hline
        \multirow{3}{*}{Condenser Lens} & 445245, Zeiss, Germany \\
         & with oil immersion attachment \\
         & 465268, Zeiss,
Germany \\
    \hline
        Dichroic Mirror & \\
    \hline
        1x telescope & Made from 2 biconvex lenses. \\
    \hline
        Quadrant Photodiode & G6849, Hamamatsu, Japan \\
    \hline
        \multirow{2}{*}{Preamplifier} & Custom Made \\
        & Öffner MSR-Technik, Germany \\
    \hline
        \multirow{2}{*}{Differential Amplifier} & Custom Made \\
        & Öffner MSR-Technik, Germany \\
    \hline
        \multirow{2}{*}{Data Acquisition Board} & $800kHz$, $16$ bit \\
         & NI 6120, National Instruments USA \\
    \hline
    \end{tabular}
    \caption[List of optical components.]{List of optical components used in the BFP microscopy setup. The components are listed in the order they appear in the beam path starting at the laser and ending with the detector and data acquisition setup.}
    \label{tab:my_label}
\end{table}

\vspace{1cm}

\section{Scattering Intensity}
For the analysis in this paper, it is important to assume that most of the beam is not scattered by the sample. This allows us to neglect the $E_{scat}E^*_{scat}$ term in Eqn. 4 as well as neglect effects of double scattering where the light scattered by one part of the sample is scattered again and contributes the light intensity at the detector. To estimate what amount of the beam is scattered by small Rayleigh scatterers, we can compute the ratio of scattered light power, $P_{scat}$ to the power of the incoming beam, $P_{beam}$. $P_{scat}$ can be calculated from the scattering cross section of a Rayleigh scattering particle (\CiteRaylCross).

\begin{align}
    P_{scat} = I_0\dfrac{8\pi}{3}\left(\dfrac{2\pi*n_s}{\lambda}\right)^4\left(\dfrac{(n_{p}/n_s)^2-1}{(n_{p}/n_s)^2+2}\right)^2 r^6
\end{align}

where $I_0$ is the intensity at the location of the Rayleigh scatterer, $n_s$ and $n_p$ are the refractive indices of the sample buffer and the particle respectively and $r$ is the radius of the particle.

The incoming beam has a Gaussian profile at the focus with a beam width of $w_0=400\ nm$ at the focus. As the Rayleigh scatterer is located at the beam focus, $I_0$ corresponds to the maximum beam intensity. We can express $P_{beam}$ as:

\begin{align}
    P_{beam} &= \underset{\textit{Infinite Plane}}{\iint} I_0 e^{-\dfrac{x^2+y^2}{w_0^2}}dxdy \\
    &= I_0\pi w_0^2
\end{align}

Therefore, the ratio of the scattering power and incoming beam power is given by:

\begin{align}
    \dfrac{P_{scat}}{P_{beam}} = \dfrac{8}{3}\left(\dfrac{2\pi n_s}{\lambda}\right)^4\left(\dfrac{(n_{p}/n_s)^2-1}{(n_{p}/n_s)^2+2}\right)^2 \dfrac{r^6}{w_0^2}
\end{align}

In the case of $110\ nm$ diameter polystyrene particles which we use as a reference in this study, $\dfrac{P_{scat}}{P_{beam}} \approx 2\times 10^{-5}$. The bead would have to exceed a diameter of $300\ nm$ before the scattering intensity would surpass $1\%$ of the beam intensity. At this size scale, the particle is experiencing an inhomogeneous electric field magnitude and more complicated effects have to be taken into account discussed further in suppl. sec. S6). We conclude that the intensity of scattered light for the structures imaged here is very small compared to the intensity of the incoming beam.

\vspace{1cm}
\section{Rayleigh Scattering Computation of $S_x$, $S_y$, and $S_z$ Signals for a Single Small Particle}
From Eqn. 4 in the main text of the paper, we see that the change in the light intensity at the location of the QPD  due to the presence of a light-scattering object is given by:
\begin{align}
    \delta I = \epsilon_s c_s \mathbb{R}e\Big\{E_{scat}E_{beam}^*\Big\}
\end{align}
In order to compute the term $\mathbb{R}e\Big\{E_{scat}E_{beam}^*\Big\}$, it is helpful to assume that the incoming laser beam is a Gaussian beam in the paraxial approximation. The magnitude of the unscattered electric field from a Gaussian beam far after the focus at position $\Vec{r}=(r,\theta, \phi)$ for polar angle $\theta$ taken with respect to the beam axis and azimuthal angle $\phi$ can be expressed approximately as:
\begin{align}
    E_{beam, \infty}(\Vec{r}) = -i \dfrac{k w_0 I^{1/2}_{tot}}{r(\pi\epsilon_s c_s)^{1/2}} \exp{\left(ikr-k^2 w_0^2 \theta^2/4\right)}
\end{align}
Here, $\epsilon_s$, $n_s$, and $c_s$ are the permittivity, refractive index, and speed of light in the sample fluid, $k=2\pi n_s / \lambda_0$ is the wave vector, $w_0$ is the beam waist, and $I_{tot}$ is the total power in the beam (\CiteGittes). Note, the factor of $-i$ represents the phase anomaly also known as the Gouy phase (\CitePralleEL, \CiteRohrbach).

For particles with diameters significantly smaller than the wavelength of the beam ($d_{part}<<\lambda$) we assume that the entire particle experiences the same electric field. We use the Rayleigh approximation to find the scattered field $E_{scat}$ from a particle displaced from the beam focal point by the vector $\Vec{x}$ to be:
\begin{align}
    E_{scat} &= \dfrac{k^2\alpha}{r}E_{Gauss}(\Vec{x})\exp{\left(ik|\Vec{r}-\Vec{x}|\right)} \\
    &\approx \dfrac{k^2\alpha}{r}E_{Gauss}(\Vec{x}) \exp{\left(ik \left( r - \dfrac{\Vec{r}\cdot \Vec{x}}{|\Vec{r}|}\right)\right)}
\end{align}
where $E_{Gauss}(\Vec{x})$ is the electric field at the particle's center and $\alpha$ is the uniform-field susceptibility given by 
\begin{align}
    \alpha=\left(\dfrac{d_{part}}{2}\right)^3\dfrac{(n_{part}/n_s)^2-1}{(n_{part}/n_s)^2+2}
\end{align}
Here $n_{part}$ and $n_s$ are the indices of refraction of the particle and of the surrounding solution respectively. We can write a Gaussian beam's electric field as (\CiteLasersSvelto,\CiteLasersPaschotta):

\begin{align}
    E_{Gauss}(\Vec{x}) &= \dfrac{2I_{tot}^{1/2}}{w_0(\pi \epsilon_0 c_s)^{1/2}} \left( \dfrac{w_0}{w(z)} \right) \\
    &\qquad\times\exp{\left(-(x^2+y^2)/w(z)^2 + ik \left(z + \dfrac{(x^2+y^2)}{2R(z)} \right) - i\phi_{Gouy}\right)} \notag
\end{align}

In the above expression, $w(z) = w_0 \sqrt{1+\left( z / z_R \right)^2 }$, $z_R = \pi w_0^2 n / \lambda$ is known as the Rayleigh range, and $\phi_{Gouy} = \arctan(z/z_R)$ is the Gouy phase shift. The terms $ikz$ and $ik(x^2+y^2)/(2R(z))$, where $R(z) = z(1+z_R^2/z^2)$ is the curvature of the wavefront, give the phase shift relative to the phase at the beam waist due to a translation parallel and perpendicular to the propagation directions of the beam, respectively.

We can now compute the change in the light intensity generated by one particle in the beam at the principal surface of the condenser lens.

\begin{align}
    \delta I &\approx \epsilon_s c_s\mathbb{R}e\Big\{E_{scat}E_{beam}^*\Big\} \\
    \begin{split}
    &\approx \dfrac{2 I_{tot} k^3 \alpha}{r^2 \pi } \left( \dfrac{w_0}{w(z)}\right) \exp \left( \dfrac{-k^2 w_0^2 \theta^2}{4} \right) \exp \left( \dfrac{-(x^2+y^2)}{w(z)^2} \right) \\
        &\hspace{2cm} \times \exp \left(  ik \left( -\dfrac{\Vec{r}\cdot\Vec{x}}{|\Vec{r}|} +z+\dfrac{x^2+y^2}{2R(z)} \right)  + \dfrac{\pi}{2} - \phi_{Gouy}\right)
    \end{split}
\end{align}

In order to find the QPD $S_x$, $S_y$ and $S_z$ signals, one has to find the total amount of light that hits each of the quadrants of the QPD. As the condenser lens maps light rays hitting its principal surface at a position $\Vec{r}=(r,\theta,\phi)$ to a lateral displacement on the QPD of $\Big( r\sin(\theta)\cos(\phi), r\sin(\theta)\sin(\phi) \Big)$, we can find the change in intensity hitting each QPD quadrant by performing the following calculation:

\begin{align}
    \delta I_{QPD,\ N} = \overset{\Theta}{\underset{\theta = 0}{\int}} \overset{N\dfrac{\pi}{2}}{\underset{\phi = (N-1)\dfrac{\pi}{2}}{\int}} \delta I(\Vec{x}) \ \ d\phi d\theta
\end{align}

where $N$ is the number of the quadrant ($I,II,III$ or $IV$), and $\Theta$ is determined by either the numerical aperture of the condenser lens or the radius of the QPD, depending on which restricts the measured light the most.

Now the $S_x$, $S_y$, and $S_z$ responses are computed as:

\begin{align}
    S_x &= \Big(\delta I_{QPD,\ I} + \delta I_{QPD,\ II}\Big) - \Big(\delta I_{QPD,\ III} + \delta I_{QPD,\ IV}\Big) \\
    S_y &= \Big(\delta I_{QPD,\ I} + \delta I_{QPD,\ IV}\Big) - \Big(\delta I_{QPD,\ II} + \delta I_{QPD,\ III}\Big) \\
    S_z &= \delta I_{QPD,\ I} + \delta I_{QPD,\ II} + \delta I_{QPD,\ III} + \delta I_{QPD,\ IV}
\end{align}

In this way, we are able to compute the expected QPD response for a dielectric particle with a displacement $\Vec{x}$ relative to the focus of the Gaussian laser beam. These QPD responses are depicted in the main text of the paper in Fig. 2a.

\vspace{1cm}
\section{Magnitude of Trapping Forces}
In order to scan the sample without disturbing its structure, we must be certain that our laser does not exert a significant force on the scanned object. For small particles, we can compute this force using a dipole approximation. When a dielectric particle is placed in the electric field produced by the laser, a dipole is induced inside the particle. As the resonance frequency of an electric dipole created due to electric polarization is very high, the dipole resonates in phase with the oscillating electric field up to low UV ($\approx 10^{15} Hz$) frequencies  (\CiteDielectDisp). One side of the dipole will feel a stronger force than the other, due the field's gradient, which results in a net force on the particle known as the gradient force. For a spherical particle, this force can be written as (\CiteGradForce):

\begin{align}
    \Vec{F}_{\nabla} &= \pi \epsilon_0 n_{s}^2 \left(\dfrac{d_{p}}{2}\right)^3 \dfrac{(n_{p}/n_s)^2-1}{(n_{p}/n_s)^2+2} \nabla |E(\Vec{r})|^2
\end{align}

We can simplify the calculation by restricting ourselves to the focal plane of the laser, which produces the largest trapping forces due to large gradients of the electric field. In the focal plane ($z=0$), the electric field of the Gaussian beam can be written as:

\begin{align}
    E_{Gauss,\ z=0} = \dfrac{2I_{tot}^{1/2}}{w_0(\pi \epsilon_0 c_s)^{1/2}}\exp\left(\dfrac{-r^2}{w_0^2}\right)
\end{align}
where $r$ is the distance from the focus, $I_{tot}$ is the power of the beam and $w_0$ is the beam waist. We can now compute the trapping force as a function of the distance from the focus.

The potential energy landscape $U(\Vec{r})$ for a small dielectric particle generated by the laser can be calculated as the path integral from some location satisfying $|\Vec{s}|=\infty$ to the location $\Vec{s}=\Vec{r}$:

\begin{align}
    U(\Vec{r}) &= \overset{\Vec{r}}{\underset{|\Vec{s}|=\infty}{\int}}-\Vec{F}_\nabla(\Vec{s})\cdot d\Vec{s} \\
    &= -I_{tot} \dfrac{n_s^2 d_{part}^3 }{2w_0^2 c_s}\dfrac{(n_{part}/n_s)^2-1}{(n_{part}/n_s)^2+2} \exp\left(\dfrac{-2r^2}{w_0^2}\right)
\end{align}

Since the power of our laser coming out of the objective lens is $I_{tot} = 0.13mW$, we can compute the potential energy as a function of displacement from the center of the trap in the $x-$direction. Similarly, we can model the filament as a line of individual particles, each experiencing their own trapping potential. Using this model, we compute the potential energy as a function of distance of a filament from the trap center as it is displaced by parallel translation in the focal plane. In the Rayleigh scattering approximation, the gradient force, $\Vec{F}_\nabla$, does not depend on the shape of the object but rather on the amount of material present. We can therefore calculate the net gradient force experienced by a filament in an optical trap, $\Vec{F}_{\nabla,\ f}$, by summing the $x$ component of the gradient force experienced by each section of an imaged filament. By symmetry, the $y$ component of the total gradient force on the fibril is 0. This gives the following result:

\begin{align}
    \Vec{F}_{\nabla,\ f}(x) &=\overset{y=\infty}{\underset{y=\infty}{\int}}\pi \epsilon_0 n_{s}^2 \left(\dfrac{d_{f}}{2}\right)^2 \textcolor{red}{\left(\dfrac{3}{2}\right)} \dfrac{(n_{f}/n_s)^2-1}{(n_{f}/n_s)^2+2} \nabla |E(x,y,z=0)|^2 dy
\end{align}
In the equation above, $n_{f}$ represents the index of refraction of collagen, and the $(3/2)$ term colored in red results from the ratio of volumes of a sphere of radius $d/2$ to a cylinder of radius $d/2$ and height $d$. We now compute the potential energy for a fibril oriented in the $y$ direction and displaced via parallel transport in the $x$ direction, $U_{f}(x)$.
\begin{align}
    U_{f}(x) &= \overset{\Vec{x}}{\underset{s=\infty}{\int}}-\Vec{F}_\nabla(s)\cdot ds \\
    &=-I_{tot} \dfrac{n_s^2}{2 w_0^2 c_s}\dfrac{(n_{f}/n_s)^2-1}{(n_{f}/n_s)^2+2}\left(\dfrac{3}{2}\right) d^2 \overset{y=\infty}{\underset{y=-\infty}{\int}} \exp\left(\dfrac{-2(x^2+y^2)}{w_0^2}\right)dy \\
    &= -I_{tot} \dfrac{3n_s^2}{w_0 c_s}\dfrac{(n_{f}/n_s)^2-1}{(n_{f}/n_s)^2+2}  d^2 \sqrt{\dfrac{\pi}{2}}w_0\exp\left(\dfrac{-2x^2}{w_0^2}\right)
\end{align}

Using the index of refraction of water ($n_s=1.333$), the index of refraction of collagen $n_{f}=1.414$ (\CiteRefrIndLeonard) and a beam waist of $w_0=400nm$, we can compute the potential energy for our laser power of $I_{tot} = 0.13mW$ (Fig. S3\iffalse\ref{PotEtrap}\fi). Note that even up to a $150\ nm$ fibril diameters, the trapping potential remains smaller than $1\ k_BT$. As the majority of experimentally observed fibrils have a diameter less than $100\ nm$, the gradient force has a smaller effect than thermal fluctuation of the filament.

\begin{figure}[htbp]
\centering\includegraphics[width=8cm]{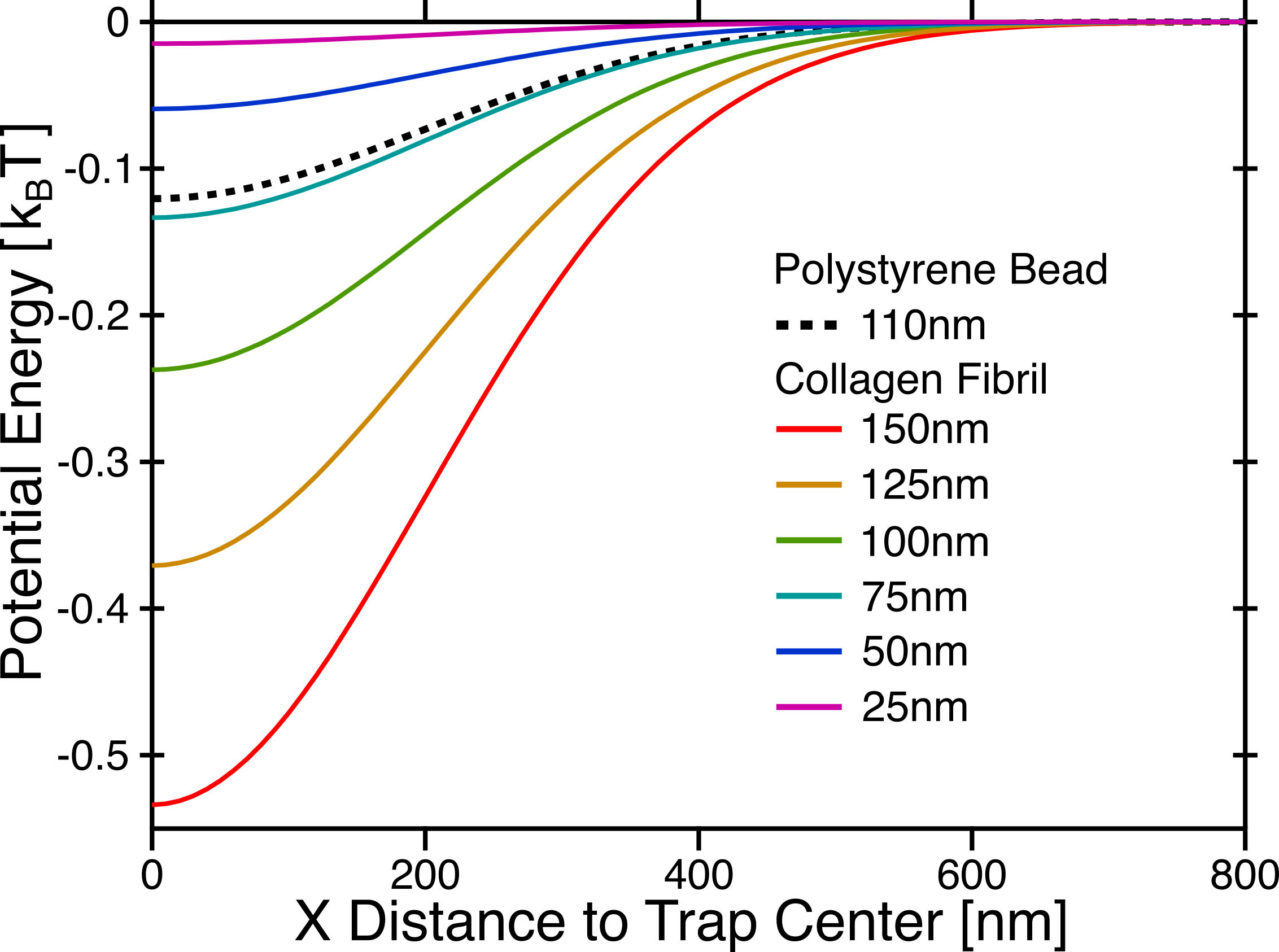}
\caption{Potential energy as a function of $x$ displacement within the focal plane for a polystyrene particle with a diameter of $110\ nm$ and for collagen fibrils with diameters varying between $25\ nm$ and $150\ nm$ oriented in the $y$ direction. For this computation, $w_0 = 400\ nm$ $n_s=1.333$, $n_{fibr}=1.414$, $I_{tot}=0.13\ mW$, and $c_s= 2.998\times 10^8/n_s\ m/s$.}
\label{PotEtrap}
\end{figure}

\vspace{1cm}

\section{Locating the Correct z-Position for Scanning}
In order to ensure that the imaging plane matches the height of the imaged object (either bead, collagen fibril, or microtubule), we scanned in either the $XZ$ or the $YZ$ plane. The average scan of a $110\ nm$ diameter bead in the $XZ$ plane is shown in Fig. S4\iffalse\ref{focPlane}\fi (a). When the laser's focal plane matches the height of the scanned object, the object experiences the largest magnitude of electric field, resulting in the largest maximum sensitivity value. To measure the response intensity, we take the directional derivative of the signal in the $x-$direction to compute the sensitivity of our detector at each location (Fig. S4\iffalse\ref{focPlane}\fi (b)). By locating the maximum sensitivity of each row (Fig. S4\iffalse\ref{focPlane}\fi (c)), we can find the height of the maximum response. Our response intensity stays within $2\%$ of the maximum value as long as the height of the imaging $XY-$plane is within $200\ nm$ of the optimal height. This also means that when measuring fibrils of different diameter on the surface, we do not have to recalculate the optimal height to scan each fibril.

\begin{figure}[htbp]
\centering\includegraphics[width=10cm]{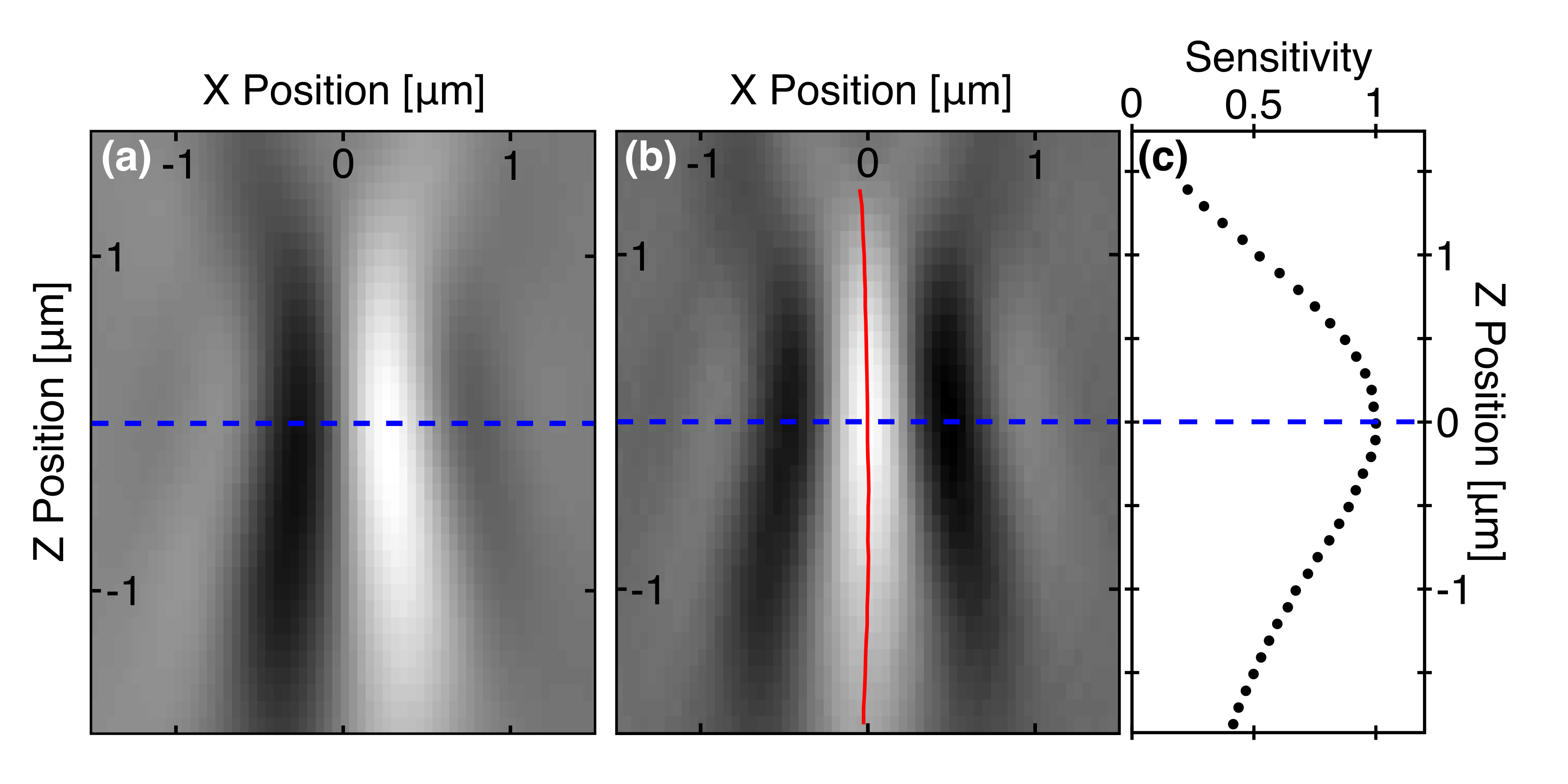}
\caption{Algorithm for locating optimal height for scan. \textbf{(a)} The $S_x$ signal for a $110\ nm$ diameter bead scanned in the $XZ$ plane. \textbf{(b)} Sensitivity of the $S_x$ signal calculated from (a). The location of maximum sensitivity for each $z$ position is found by fitting a parabola around the maximum sensitivity value of each row. The location for the maximum sensitivity of each row is shown in red. \textbf{(c)} The maximum sensitivity at each $z$ position plotted as a function of $z$ position. By fitting a parabola around the maximum value, the optimal $z$ position for the scan in the $XY$ plane is found. Optimal $z$ position is shown in blue}
\label{focPlane}
\end{figure}

\section{Rayleigh Scattering Computation for Large Beads}
The Rayleigh scattering computation described in Section S2 of the supplemental materials assumes that the magnitude of the laser's electric field is constant throughout the bead leading to the response magnitude scaling with bead volume. This assumption does not hold true for beads larger than the beam waist. From Section 3 of the paper, we conclude that the response to a scan of multiple beads can be approximated by the sum of multiple single bead responses with respective shifts. Therefore, to construct the response of a single large bead, we model it as consisting of many small individual beads. We impose the constraint that the total volume of the small beads match the volume of the large bead. Each individual bead's response can now be scaled proportionally to the magnitude of the electric field. By computing the weighted sum of the responses of all individual beads with the associated shifts and weights corresponding to the magnitude of the electric field at each beads location, we can construct the response of a large bead.\\
\\
This computational method shows a deviation from the scaling law predicted by the computation in Section S2 as one approaches larger bead sizes. The Rayleigh scattering computation in Section S2 is limited by two assumptions: first that the particle is significantly smaller than the beam waist, and therefore the laser's electric field has the same magnitude at all particle locations; second that the scattered light is negligible relative to the intensity of the unscattered beam, allowing us to neglect interactions between the bead and the scattered light. Using the improved method of computing the Rayleigh scattering response of the particle, we only have to make the second assumption. By comparing the Rayleigh and Mie scattering computations, we see that the Rayleigh scattering computation is now accurate for a much larger range of bead sizes (see Fig. 3 in the main text). 

\vspace{1cm}
\newpage
\section{Multiple Bead Response}
\ 
\begin{figure}[htbp]
\centering\includegraphics[width=10cm]{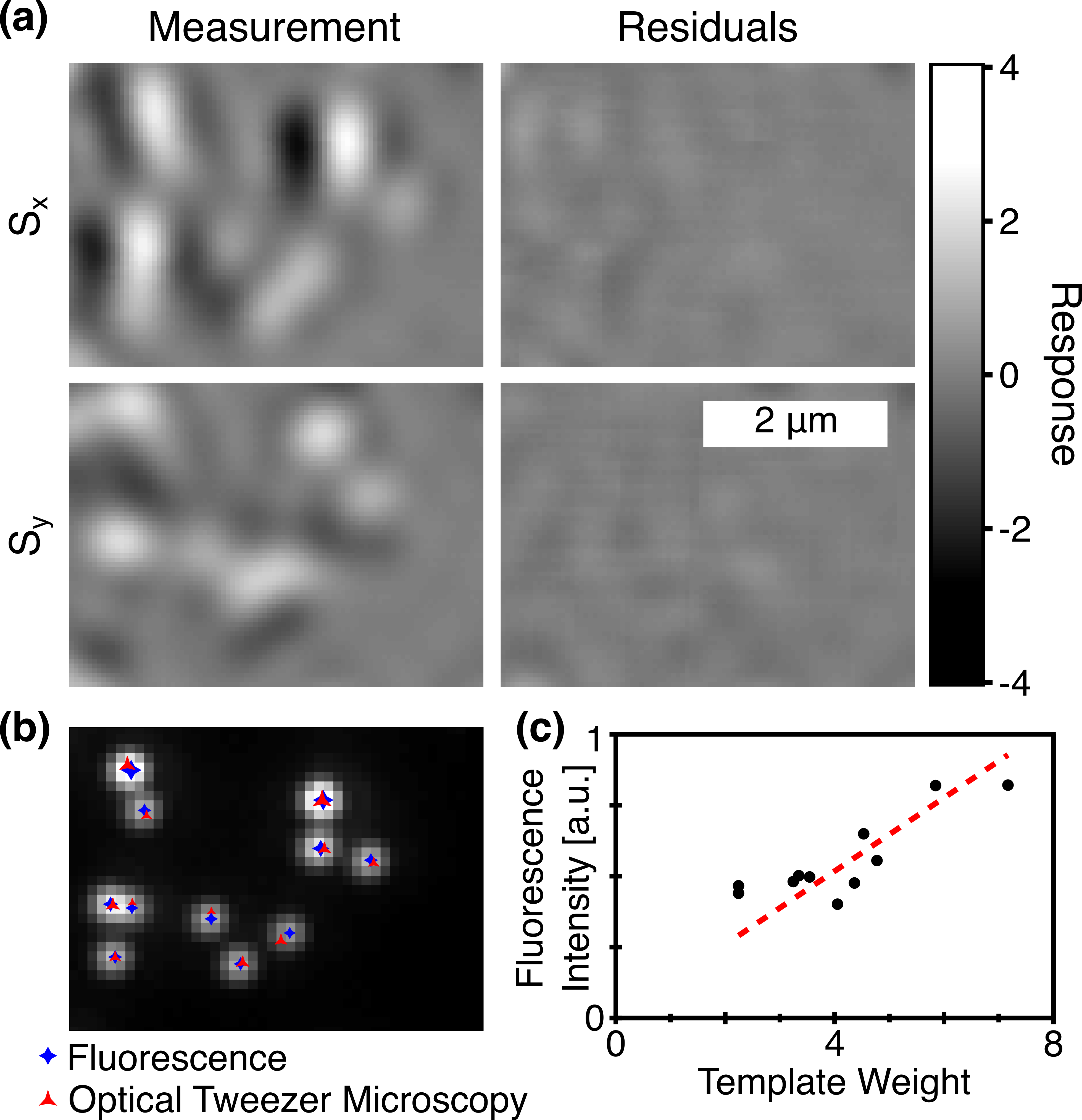}
\caption{Images of multiple nearby beads compared to their reconstruction via the sum of individual bead signals. \textbf{(a)} $S_x$ and $S_y$ signals for multiple $170\ nm$ diameter beads on the left. After fitting a sum of 11 single bead responses, the residuals are plotted on the right. Responses in this figure are rescaled such that $\left(\partial S_x/\partial x\right)_{max}=1$ for a $110\ nm$ polystyrene bead. \textbf{(b)} Fluorescence image of the beads with bead locations computed by fitting Gaussian profile to the fluorescence image and by using the custom template fit function (main text Eqn. 9). The size of the marker is scaled proportionally to the amplitude found from the fluorescence or template fit. \textbf(c) Scaling of fluorescence intensity in arbitrary units with weight from template fit. The weight found from the template fit is scaled such that a single $110\ nm$ particle would have the weight of $1$.}
\label{rm11Beads}
\end{figure}

\section{Vibration and Intensity Fluctuations Noise}

As the $S_x$ and $S_y$ signals are differential measurements, these signals are susceptible to vibrational noise as well as laser pointing noise. Our optical setup was isolated from external vibrations using a pneumatic support system. Air currents, as well as external light was minimized by covering the beamline with black poster-board. The full setup was also covered with a plastic curtain. This further reduced air currents and prevented dust from entering the optical setup. The relative contributions of laser power noise and laser pointing noise can be quantified by passing the beam through a water-filled sample and looking at the noise in the resulting QPD $S_x$ and $S_z$ signals. As the $S_z$ signal collects the full transmitted light intensity and the beam is not clipped by the edges of the QPD, $S_z$ is sensitive to laser power and not sensitive to laser pointing fluctuations and system vibrations. However, as $S_x$ is a differential signal, it is sensitive to fluctuations and vibrations that displace the beam, but is insensitive to laser power fluctuations when correctly centered. The variance in $S_x$ is $2.5$ times smaller than the variance in $S_z$ signal implying that fluctuations in direct beam's electric field strength are a larger contribution to the noise. The efforts to reduce vibrations and air currents are working. The effect of laser power noise, pointing noise, and vibrations can be further reduced by increasing the integration time. However, at long integration time, stage drift effects become more significant.